\documentclass[a4paper]{article}

\pdfoutput=1

\usepackage{graphicx}
\usepackage{psfrag}
\usepackage[small]{caption} 
\usepackage{amsmath}
\usepackage{amsfonts}
\usepackage{amsthm}
\usepackage{hyperref}
\usepackage{subfigure}
\usepackage[mediumspace,mediumqspace,squaren]{SIunits}

\allowdisplaybreaks[4]
\numberwithin{equation}{section}
\numberwithin{figure}{section}

\newcommand{\naturali}{{\mathbb{N}}}

\renewcommand{\epsilon}{\varepsilon}
\renewcommand{\phi}{\varphi}
\renewcommand{\theta}{\vartheta}

\renewcommand{\d}[1]{\mathinner{\mathrm{d}{#1}}}

\title{A hybrid model for large neural network description}

\author{Anna Cattani\\ \small Center for Neuroscience and Cognitive Systems@UniTn,\\ \small Istituto Italiano di Tecnologia, Rovereto, Italy\and Sergio Solinas \\ \small Department of Brain and Behavioural Science, University of Pavia, Pavia, Italy\and Claudio Canuto \\ \small Department of Mathematical Sciences, Polytechnic University of Turin, Italy}

\begin{document}

\maketitle

\begin{abstract}
The aim of the present paper is to efficiently describe the membrane potential dynamics of neural populations formed by species having a high density difference in specific brain areas. We propose a hybrid model whose main ingredients are a conductance-based model (ODE system) and its continuous counterpart (PDE system) obtained through a limit process in which the number of neurons confined in a bounded region of the brain is sent to infinity. Specifically, in the discrete model each cell of the low-density populations is individually described by a set of time-dependent variables, whereas in the continuum model the high-density populations are described as a whole by a small set of continuous variables depending on space and time. Communications among populations, which translate into interactions among the discrete and the continuous models, are the essence of the hybrid model we present here. Such an approach has been validated reconstructing the ensemble activity of the granular layer network of the Cerebellum, leading to a computational cost reduction. The hybrid model reproduced interesting dynamics such as local microcircuit synchronization, travelling waves, center-surround and time-windowing. 
\end{abstract}

  \smallskip

  \noindent \footnotesize{ \textit{2010 Mathematics Subject Classication.} Primary: 34C60, 35K57; Secondary: 92C42, 05C90}
  
    \smallskip
    
  
  \smallskip
    
  \noindent \footnotesize{ \textit{Contact:} Anna Cattani: anna.cattani@iit.it; Sergio Solinas: smgsolinas@gmail.com; Claudio Canuto: claudio.canuto@polito.it.}

\normalsize
\section{Introduction}
Interesting phenomena in the brain often involve complex networks with an extremely large number of neurons. The description at the microscopic level of the whole network, i.e., the modelling of each single neuron and synapse, would lead to numerical models of prohibitive computational cost, even on the most advanced computers. 
The difficulties of such a description may be alleviated to some extent by identifying a hierarchy among interacting populations of neurons, and by using models with different resolution and cost for simulating the behaviour of different  populations.
Cell density may be a criterion to identify families of neurons and to partition the network in a multi-level manner, where each level corresponds to one or more species with comparable density.
In the simplest situation of a two-level organization, this option leads to describe each neuron of the low-density population(s) by means of an ODE system, and to characterize the high-density population(s) by exploiting a PDE system that describes the family as a continuum. The hybrid model collects the ODE and the PDE systems, as well as the fundamental interactions among them.
Several efforts, which have resulted in the formalisation of different models, have been made to understand and reproduce the activity of high-density populations by reducing the degrees of freedom from many, i.e., the variable states for each neuron, to few. \textit{Mean-field}, \textit{neural mass} and \textit{neural-field models} are some of the results of various ``passage to the continuum'' approaches. A review concerning these models can be found in \cite{Bressloff}-\cite{Deco}. The major difference between neural-field models - such as the one we are going to present - and the others lies in the fact that the formers account for the spatio-temporal evolution of the variables, rather than considering just the temporal evolution of them. A pillar formalisation of a neural-field model is proposed in \cite{Amari1977}-\cite{WilsonCowan}-\cite{WilsonCowan2}, in which the macroscopic state variable is the mean firing rate. A more general neural-field model, not necessarily involving only firing rate variables, is presented in \cite{Touboul}.
 
We obtain a continuum model for the action potential of a dense population of neurons by starting from a discrete model and letting the number of neurons tend to infinity while keeping them confined in a bounded region. We identify limit operators, acting on the continuous variables, describing specific interactions: in particular, electrical couplings (``gap junctions'') are modelled in the limit by the Laplace differential operator, as it has been rigorously justified in \cite{CanutoCattani}; on the contrary, chemical synaptic couplings produce non-local integral operators, i.e., spatial convolutions with suitable kernels (see e.g. Sect. 9.2 in \cite{Ermentrout}). Once the expressions of both the discrete and the continuum model have been set, we describe in a fairly general form how the two models reciprocally interact, producing a hybrid model: aside of terms in the equations describing interactions between ``homogeneous''  (i.e., discrete-discrete, or  continuous-continuous) variables,  new terms are added to account for the ``heterogeneous'' interactions (i.e., between discrete and continuous, or  continuous and discrete, variables).

To validate our new method in a complete workflow we applied it to a realistic computational problem, the reconstruction of the Cerebellum granular layer network (GLN). This  brain area shows a simple network structure yet capable of generating complex activity patterns.
This network layer is densely populate by granule cells (GrCs) and sparsely by Golgi cells (GoCs) providing an optimal application for our modeling approach.
The proposed hybrid model was specialized to the description of the interactions between GrC and GoC populations in the Cerebellum. Interesting dynamics such as local microcircuit synchronization,  center-surround and time-windowing, as already described in a previous and more biologically detailed model \cite{Solinas2010}, are reproduced by the proposed model. Moreover, our model show the emergence of travelling waves of network activity elicited by a specific input configuration.

\section{Materials and Methods}
\subsection{The hybrid model}

In this section, in order to introduce the hybrid model, we first show how to model each individual neuron belonging to the same population. Here, intra-population communications are taken into account. Secondly, due to the fact that the number of neurons even in a small brain area is often huge, we perform a continuum limit of the discrete model that describes single neurons, obtaining a continuous model. Finally, we present the hybrid model in which the discrete and the continuous models interact with each other.

Let us start by analysing how to describe the dynamics of each individual neuron $i$ in the network, with $i=1,\cdots N$, where $N$ is the number of neurons in a population. Precisely, we consider three variables: the voltage-like variable $v_i$, the recovery variable $r_i$, and the $s_i$ variable which describes the fraction of open channels in the synapses. In the most general fashion, each neuron is influenced by other neurons in the network by means of electrical and chemical synapses, and its dynamics is also driven by terms that describe the basic properties of neural excitability. All these ingredients are taken into account in the following general model:
\begin{equation}
\begin{aligned}
\frac{\d v_i}{\d t}&=f(v_i,r_i)+I_{\rm{gap}}^i+I_{\rm{syn}}^i\;,\\
\frac{\d r_i}{\d t}&=g(v_i,r_i)\;,\\
\frac{\d s_i}{\d t}&=\alpha_i(1-s_i)H_\infty (v_i-v_T)-\beta_i s_i\;,
\end{aligned}
\label{Eq:DiscCompleteModel}
\end{equation}
where, $I_{\rm{gap}}^i$ is the input current that accounts for electrical synapses, and $I_{\rm{syn}}^i$ is that for chemical synapses. In particular,
\begin{equation}
\begin{aligned} 
I_{\rm{gap}}^i&=d\sum_{j\in\mathcal{Q}(i)}(v_{j}-v_{i})\;,\\
I_{\rm{syn}}^i&=g_{\rm{syn},\textit{i}} \sum_{j\in\mathcal{B}(i)} w_{ij}s_{j}(v_i-v_{\rm{syn},\textit{j}})\,,
\end{aligned}
\end{equation}
where $\mathcal{Q}(i)$ and $\mathcal{B}(i)$, resp., collect the indexes of neurons connected to the $i$-th one by means of electrical and chemical synapses, resp., $w_{ij}$ are positive weights describing the directed connection strength from $j$ to $i$, $d>0$ is the diffusion coefficient, $g_{\rm{syn},\textit{i}}>0$ is the synaptic efficacy, and $v_{\rm{syn},\textit{j}}$ is the reversal potential of the presynaptic neuron whose sign determines the synapse nature, either excitatory or inhibitory. In \cite{Destexhe} and \cite{Ermentrout}, a detailed classification of synaptic reversal potentials, linked to distinct neurotransmitter/receptor pairs, is specified. 
Furthermore, among the wide variety of models which describe the basic properties of neural excitability, we select the FitzHugh-Nagumo model \cite{FitzHugh} phenomenologically extracted from the biophysically-based Hodgkin-Huxley model. Thus,
\begin{equation}
\begin{aligned}
f(v_i,r_i)&=-v_i(a-v_i)(1-v_i)-r_i\;,\\
g(v_i,r_i)&=bv_i-cr_i\;.\\
\end{aligned}
\end{equation}
Here, $a,\,b,\,c\in \mathbb{R}^+$ are parameters chosen so that $v_i$ is a fast variable and $r_i$ is a slow one.
Finally, in the third equation in \eqref{Eq:DiscCompleteModel}, $\alpha$ and $\beta$ are positive parameters describing the forward and backward rate constants for transmitters binding, $v_T$ is an \textit{a priori} fixed threshold, and $H_\infty=H_\infty(z)$ is the Heaviside function such that $H_\infty=0$ if $z<0$ and $H_\infty=1$ otherwise. 
The model \eqref{Eq:DiscCompleteModel} obviously should be supplemented by suitable initial conditions for the variables $(v_i,r_i,s_i)$. 

In order to avoid prohibitive computational costs when the density of cells in a population is too high, we perform a ``passage to the limit'' as the number of neurons $N$ tends to infinity in \eqref{Eq:DiscCompleteModel}. In this way, we 
capture the dynamics of a neuronal population as a whole by describing three continuous variables $v(x,t)$, $r(x,t)$ and $s(x,t)$ (having the same meaning as in \eqref{Eq:DiscCompleteModel}), where $x$ is the spatial variable. Specifically, in the limit case of $N\rightarrow\infty$ in a fixed and bounded spatial region $\Omega\subset\mathbb{R}^m$, with $m\in\{1,2,3\}$, the discrete model \eqref{Eq:DiscCompleteModel} leads to the following integro-differential system of equations (in which the $t$-dependence of each variable is ignored for simplifying notation):
\begin{equation}
\label{Eq:ContCompleteModel}
\begin{aligned}
\frac{\partial v}{\partial t}(x)&=f(v(x),r(x))+d^\ast\Delta v(x)-g_{\rm{syn}}\int_{\mathcal{R}(x)} w(x,y)s(y)(v(x)-v_{\rm{syn}}(y)) \rm{d} y && \\
\frac{\partial r}{\partial t}(x)&=g(v(x),r(x))&& \\
\frac{\partial s}{\partial t}(x)&=\alpha(1-s(x))H_\infty (v(x)-v_T)-\beta s(x)\;,
\end{aligned}
\end{equation}
supplemented by boundary conditions for $v$ and initial conditions for $v$, $r$, $s$. Here, $d^\ast$ is the diffusion coefficient, $g_{\rm{syn}}>0$ is the synaptic efficacy, and $\mathcal{R}(x)$ denotes a region centered in $x$. The whole electrical synapse term, i.e.  $d^\ast\Delta v(x)$, is the result of two equivalent methods that lead to a non-trivial continuum limit, as shown in \cite{CanutoCattani}. On the other hand, the integral form of the chemical synapse term, i.e. $g_{\rm{syn}}\left(\int_{\mathcal{R}(x)} w(x,y)s(y)(v(x)-v_{\rm{syn}}(y)) \rm{d} y\right)$, is due to the fact that the set $\mathcal{B}(i)$ in \eqref{Eq:DiscCompleteModel} does not shrink to a point as $N \to \infty$, as explained in \cite{Ermentrout} and \cite{TesiDott}. Furthermore, we refer to \cite{TesiDott} for a discussion on the mathematical well-posedness of this model. Afterwards, in order to distinguish between the discrete and the continuum systems, variables in the continuum configuration will be indicated by Greek letters.

As already mentioned in the introduction, by comparing the cell densities we may diversify the description of the populations in the network. Specifically, this comparison determines if a population is described by a set of discrete systems or by a continuous model.
However, the key point is that neurons are linked to each others in a very intricate fashion depending on the brain areas. It follows that signal transmission among populations, in addition to intra-populations connectivity, is an important feature to be taken into account to explore the emergent network dynamics. The essence of the hybrid model lies in the interaction coupling terms among different populations.

By considering for simplicity two populations only, on the one hand the set of cells in the low-density population is described by an ODE system:
\begin{equation}\begin{aligned}
\frac{\d v_i}{\d t}&=f(v_i,r_i)+\phi(v_i;v_j,s_j)+\Phi(v_i;\omega,\sigma)+I^i_{\rm{ext}}\;,\\
\frac{\d r_i}{\d t}&=g(v_i,r_i)\;,\\
\frac{\d s_i}{\d t}&=\alpha_i(1-s_i)H_\infty (v_i-v_T)-\beta_i s_i\;,
\end{aligned}
\label{Eq:MultiscaleDiscrete1}
\end{equation}
where 
\begin{equation}
\label{Eq:phi}
\begin{aligned}
\phi(v_i;v_j,s_j)&=d\sum_{j\in \mathcal{Q}(i)}(v_j-v_i)-g_{\rm{syn}}\sum_{j\in\mathcal{B}(i)}w_{ij}s_j(v_i-v_{\rm{syn},\textit{j}})
\end{aligned}
\end{equation}
takes into account inputs from other cells belonging to the same low-density population, whereas 
\begin{equation}
\label{Eq:Phi}
\begin{aligned}
\Phi(v_i;\omega,\sigma)&=\delta\Delta\omega (x_i)-\gamma_{\rm{syn}}\int_{\mathcal{R}_i}w(i,y)\sigma (y) (v_i-\omega_{\rm{syn}}(y))\d y
\end{aligned}
\end{equation}
describes the signal transmission coming from the continuous population. 
Here, $x_i$ indicates the spatial position of the neuron labelled by $i$ from the discrete family, whereas $\mathcal{R}_i$ is the region occupied
by the neurons from the continuous family whose synapses influence neuron $i$. 
The term $I^i_{\rm{ext}}$ represents an external current coming from sources different from the two species here considered.
On the other hand, the high-density population is characterized by a PDE system:
\begin{equation}\begin{aligned}
\frac{\partial\omega}{\partial t}&=F(\omega,\rho)+\psi(\omega,\sigma)+\Psi(\omega;v,s)+\mathcal{I}_{\rm{ext}}\;,\\
\frac{\partial\rho}{\partial t}&=G(\omega,\rho)\;,\\
\frac{\partial\sigma}{\partial t}&=\alpha(1-\sigma)H_\infty (\omega-\omega_T)-\beta \sigma\;,
\end{aligned}
\label{Eq:MultiscaleContinuum1}
\end{equation}
where, similarly to \eqref{Eq:phi}, 
\begin{equation}
\label{Eq:psi}
\begin{aligned}
\psi(\omega,\sigma)(\xi)&=\delta\Delta\omega(\xi)-\gamma_{\rm{syn}}\int_{{\mathcal R}(\xi)}w(\xi,y)\sigma (y)(\omega(\xi) -\omega_{\rm{syn}}(y))\d y
\end{aligned}
\end{equation}
concerns interactions within the continuum population, while
\begin{equation}
\label{Eq:Psi}
\begin{aligned}
\Psi(\omega ;v,s)(\xi)&=d\sum_{j\in \mathcal{Q}(\xi)}(v_j-\omega(\xi))-g_{\rm{syn}}\sum_{j\in\mathcal{B}(\xi)}w(\xi,j)s_j(\omega(\xi)-v_{\rm{syn},\textit{j}})
\end{aligned}
\end{equation}
describes the interactions between species, and $\mathcal{I}_{\rm{ext}}=\mathcal{I}_{\rm{ext}}(\xi)$ is an external current. 
We call hybrid the model constituted by systems \eqref{Eq:MultiscaleDiscrete1}-\eqref{Eq:Phi} and \eqref{Eq:MultiscaleContinuum1}-\eqref{Eq:Psi}.

\subsection{Application to the Cerebellum granular layer network}
The formalization of the hybrid model developed above is suitable for describing a variety of networks in the brain characterized by a large difference in their population densities. Among others, the olfactory bulb, the striatum, the granular layers of the dorsal cochlear nucleus and the Cerebellum cortex are suitable to be efficiently represented with our new method. Out of these exempts the Cerebellum cortex is the most extensively studied and modelled network.

The network structure can be abstracted following previous modelling works \cite{Solinas2010code,Solinas2010} but keeping it sufficiently adeherent to the biological reality to show the versatility of the method to reproduce neural network dynamics observed in brain tissues. In the specific, the limit used to push to continuum the representation of the neuronal population with high density, could rise issues on the reproducibility of network dynamics generated by a network composed by many small, independent units yielding unwanted diffusion of the activity across the network. A case that cannot yet be excluded in the nature of the real brain tissues but that was excluded in the biologically realistic simulations \cite{Solinas2010}. Moreover, the practical test should highlight the cooperation of the discrete units with the continuum model.

Interests in Cerebellum dates back to the morphological studies carried out by Ramon y Cayal and Camillo Golgi, the electroencephalography studies carried out by Adrain \cite{adrian1935discharge} and the motor impairment manifest in World War I and II patients with cerebellar lesions studied by Holmes \cite{Holmes1917a}. Only later on, the fine Cerebellum structure inspired theories linking the network structure to a function starting, with Braitemberg \cite{Braitenberg1958}, Marr \cite{Marr}, Albus \cite{Albus} and Ito \cite{ito1984cerebellum}, a research work yet to be accomplished. Its peculiar structure comprehends series of highly regular, repeating units, each of which contains the same basic microcircuit. The similarity in repeating units, from architectural and physiological perspectives, implies that different regions perform similar computational operations on different inputs. These inputs originate from different parts of the brain, spinal cord, and sensory system projecting directly into the Cerebellum. In turn, the Cerebellum projects to all motor systems.
Despite the regularity of the Cerebellum facilitates its description, it remains a network able to generate complex dynamics whose potentialities and functionalities are not yet fully understood. 

Few cellular populations in the Cerebellum cortex compose this geometrically regular network and are localised in three well distinct layers called \textit{molecular}, \textit{Purkinje}, and \textit{granular}. 
The latter is densely populated by GrCs (density $4.000.000/mm^3$) and sparsely by GoCs. 
The key point to support the application of our new modelling method is that the number of GoCs significantly differs from that of GrCs: GoCs are very few compared to GrCs \cite{Korbo1993,Solinas2010,Billings2014} in the reason of about $1:400$. Thus, by virtue of this strong density difference, the exploitation of combined discrete and continuum models becomes interesting. In particular, the variables $(v_i,r_i,s_i)$ describe each GoC through \eqref{Eq:MultiscaleDiscrete1}, while $(\omega,\rho,\sigma)$ portray the GrC species as a whole by means of \eqref{Eq:MultiscaleContinuum1}.
We focused our test study to reproduce the transformation imposed to the input signals by the Cerebellum granular layer network (GLN).  The ultimate output of the GLN provide excitatory input through their axons in the molecular layer to the Purkinje cells which constitute the only output pathway of the cerebellum cortex. 
The GLN is composed of two main network pathways, a feedforward path and a loop or feedback path, where both Granular cells (GrCs) and Golgi cells (GoCs) receive external excitatory inputs by the Mossy fibers (MFs) originating from the precerebellar nuclei neurons. MFs excite both cell populations duplicating their input into two pathways. Along the feedback path MFs excite GrCs. These excite GoCs through the ascending axon and the parallel fibers (PFs), and GoCs, in turn, inhibit GrCs. In a compact writing:
MF-GrCs-PFs-GoCs-GrCs. The second or feedforward path is constituted by the excitatory input from MFs to GoCs which terminates inhibiting GrCs. This pathway is MF-GoCs-GrCs.

\begin{figure}
  \centering
    \subfigure{\includegraphics[width=0.4\textwidth]{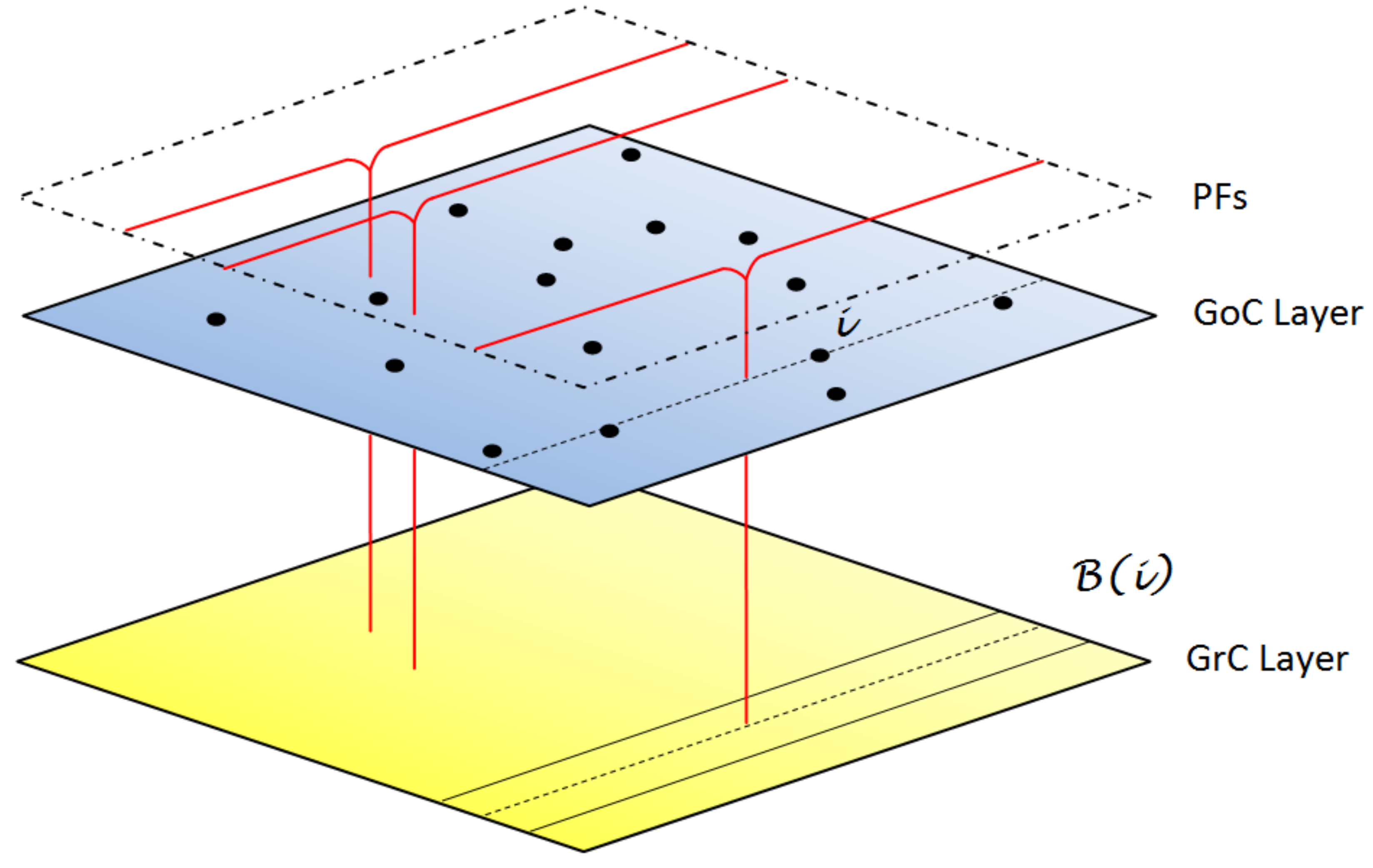}}
	  	  \subfigure{\includegraphics[width=0.4\textwidth]{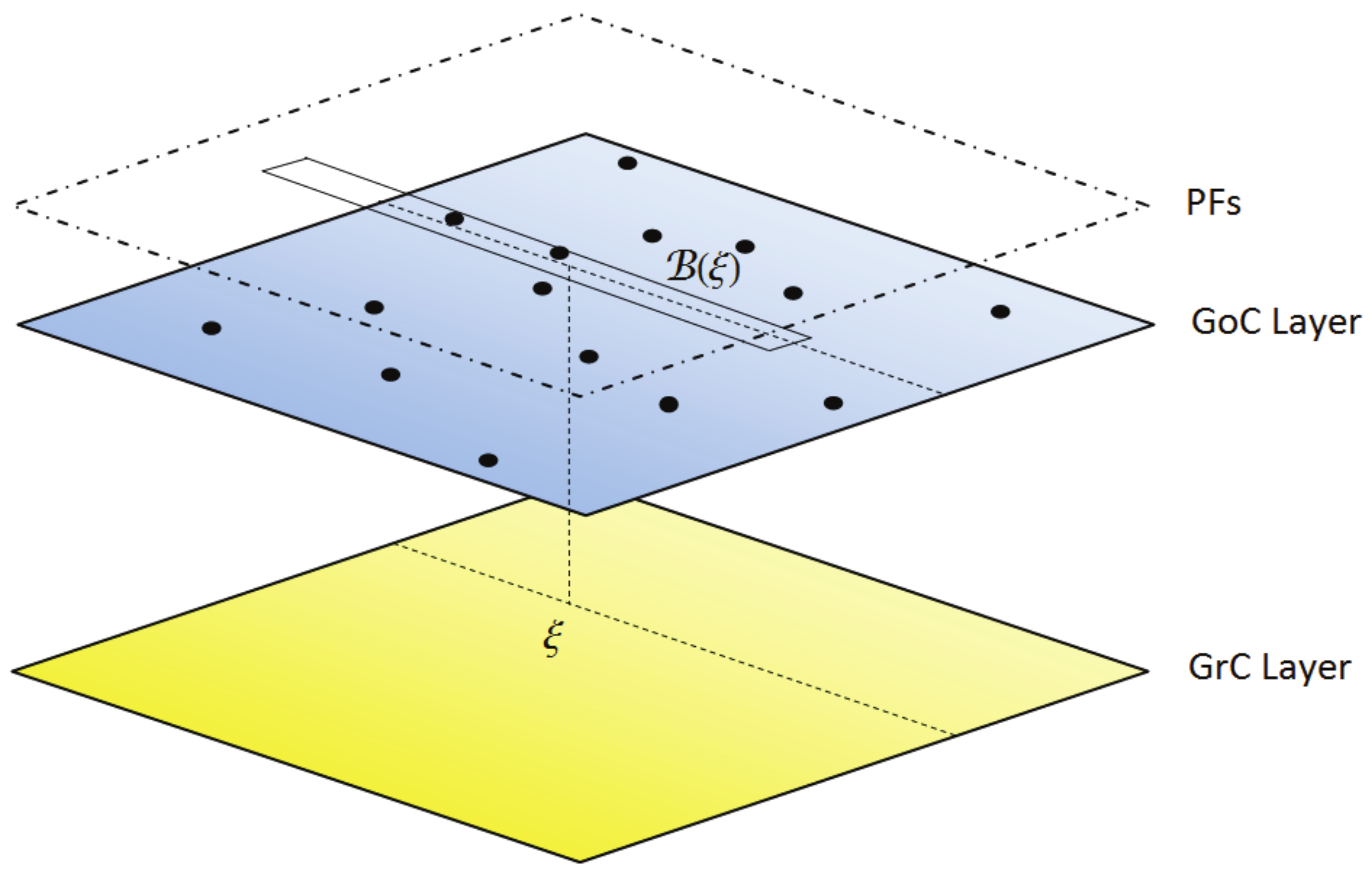}}
	  \caption{\label{Fig:LayerLinks}Connection topology between GrCs and GoCs from a postsynaptic neuron perspective: GrCs linked to the $i$-th GoC (left) and GoCs which are connected to the GrC at the point $\xi$ (right).
	  }
\end{figure}

Inspired by assumptions in \cite{DeSchutter} and for modelling purposes, we consider the two populations belonging to two-dimensional parallel layers, as described in Fig. \ref{Fig:LayerLinks}. The bottom one is constituted by the GrC continuum and the upper one collects GoCs. A third layer, above them, collects PFs. In reality, GoC somata and GrCs are located in the just mentioned \textit{granular layer} while the site where GoC dendrites receive input from the GrC axons (PFs) is in the \textit{molecular layer}.

Let us now define our model topology and connectivity in the GLN taking into account the fine structure of the biological network. The model was build to reproduce a GLN fraction with size \unit{1500}{\micro\meter} along the sagittal axis, \unit{500}{\micro\meter} on the transverse axis and \unit{100}{\micro\meter} thick. However, in our representation the thickness of this flat volume is disregarded. In our model all spatial units are normalized to the network edge length. GoCs and GrCs are assumed to belong to two rectangular domains of size $[0,3]\times[0,1]$, one on top of the other (Fig. \ref{Fig:LayerLinks}). 
The projection of MFs inside the GLN shows an abundant parasagittal branching. Each MF innervates multiple cerebellum lobules. Within the lobule, local branching gives origin to small clusters of about 8 MF terminals in a rectangular area of \unit{200}{\micro\meter} along the transverse axis and \unit{150}{\micro\meter} along the sagittal axis \cite{Sultan2003a,Solinas2010}, data from the rat cerebellum. About 50 GrCs project their dendrites (maximum length \unit{40}{\micro\meter}, mean length \unit{13.6}{\micro\meter}) on a MF terminal. Therefore, the activation of a single MF should give rise to small spots of activated GrCs with response intensity degrading from centre to periphery. In our model, the GrC population is represented as a continuos sheet split into vertices by tessellation allowing the calculation of numerical solutions. In this configuration, we assume that MF terminals provide excitatory input to a subset of the vertices. A diffusive term in the PDE spreads the input to the neighbouring vertices, the intensity fading to none for a distance equal to \unit{40}{\micro\meter}. 
GoCs receive excitatory input from MF terminals from a wider area as GoC dendrites are longer than GrC dendrites and span a larger GLN volume \cite{Dieudonne1998}. 

Each GoC arborized axon reaches the granular layer throughout a parallelepiped volume \cite{Barmack2008c} elongated along the sagittal direction, whose projection on the two-dimensional granular layer is a rectangle \unit{650}{\micro\meter} long and \unit{180}{\micro\meter} wide. A GoC sparsely inhibits GrCs lying inside the rectangle. 
GrC axons, i.e., PFs, ascend to the molecular layer, bifurcate, and run parallel to each other in either direction along the transversal axis, our $x$-axis, for a few $mm$ crossing the GoC apical dendrites. Each PF synapses onto many GoC dendrites along its path.  The GoC apical dendrites branch out in all directions sampling PF input from a cylinder in the ML represents in the original model by a circle of radius \unit{50}{\micro\meter} \cite{Dieudonne1998}. 
Therefore in our model, a GoC provide inhibitory input to all the GrCs located within a rectangle elongated along the sagittal axis, with length $1.3$ and width $1/2.8$. Each GrC influences all GoCs in a rectangle elongated along the transverse axis, covering the entire GLN extension, and narrow along the sagittal axis, covering $1/10$ on either side of the PF wide stripe of the GLN (see Fig. \ref{Fig:LayerLinks}). 
Notably, GoGs receive chemical excitatory synapses by GrCs.
Furthermore, GoCs are linked among each other by gap junctions connecting their apical dendrites \cite{Vervaeke2010}. This electrical coupling is represented in our model by a diffusion term between the vertixes of the discrete model, i.e. in a first approximation a GoC is coupled only with its nearest neighbours.

As already mentioned above, the Golgi cell system can be described by the model \eqref{Eq:MultiscaleDiscrete1}; the general expression of the functions $\phi$ and $\Phi$, given in \eqref{Eq:phi} and \eqref{Eq:Phi}, takes here the following specific form:
\begin{equation}
\begin{aligned}
\label{Eq:GoC}
\phi(v_i;v_j,s_j)&=d\sum_{j\in \mathcal{Q}(i)}(v_j-v_i)\\
\Phi(v_i;\omega,\sigma)&=-\gamma_{\rm{syn}}\int_{\mathcal{R}_i}w(i,y)\sigma (y)(v_i-\omega_{\rm{syn}})\d y\;.\\
\end{aligned}
\end{equation}
Moreover, $I^i_{\rm{ext}}=I^i_{\rm{mossy}}$ is the excitatory input due to the MFs.
Let us recall that, in \eqref{Eq:GoC}, the reversal potential $\omega_{\rm{syn}}$ may depend upon the presynaptic neurons and, thus, it must be included in the integral term. However, since only GrCs influence GoCs by means of excitatory chemical synapses, we suppose $\omega_{\rm{syn}}$ to be constant and we bring it out of the integral, obtaining
\begin{equation}
\nonumber
\Phi(v_i;\omega,\sigma)=-\gamma_{\rm{syn}}\biggl(\int_{\mathcal{R}_i}w(i,y)\sigma (y)\d y\biggr)(v_i-\omega_{\rm{syn}})\;.\\
\end{equation}
The set $\mathcal{R}_i$ determines the area containing those GrCs which synapse onto the $i$-th Golgi cell. Taking into account that GrCs excite GoCs through the PFs, as specified above, we consider $\mathcal{R}_i$ as a thin rectangle whose horizontal symmetry axis is determined by the $i$-th cell projection (see Fig. \ref{Fig:LayerLinks}, left). The rectangle area is chosen by fixing a reasonably small parasagittal extension.



Furthermore, concerning the coupling term between the two populations, it is well known that GrCs receive inhibitory chemical synapses from GoCs. Thus, the GrC continuum is described by the model \eqref{Eq:MultiscaleContinuum1}, where the functions $\psi$ and $\Psi$, introduced in \eqref{Eq:psi} and \eqref{Eq:Psi}, take the following specific form:
\begin{equation}
\label{Eq:GrC}
\begin{aligned}
\psi(\omega,\sigma)(\xi)&=\delta\Delta\omega(\xi)\;,\\
\Psi(\omega; v,s)(\xi)&=-g_{\rm{syn}}\biggl(\sum_{j\in\mathcal{B}(\xi)}w(\xi,j)s_j\biggr)(\omega (\xi)-v_{\rm{syn}})\;.
\end{aligned}
\end{equation}
As above, the reversal potential $v_{\rm{syn}}$ of presynaptic GoCs is supposed to be constant and then it is not involved in the summation. In order to consider inputs from Mossy Fibers, we set $\mathcal{I}_{\rm{ext}}=\mathcal{I}_{\rm{mossy}}$.  The discrete set $\mathcal{B}(\xi)$ collects the indexes of GoCs which influence the GrC continuum at the point $\xi$, thus describing the connection topology. According to \cite{Barmack2008c}, a GoC axon reaches a rectangular region in the granular layer, centered on its soma; therefore, a possible choice is:
\begin{equation}
\label{Eq:TopologyRect}
\mathcal{B}(\xi):=\{j\in\naturali: x_j\in R_{\xi}\}\;,
\end{equation}
where $R_{\xi}$ denotes such a rectangle centered on the projection of $\xi$ on the GoC plane and oriented perpendicularly to the $\mathcal{R}_i$ direction (see Fig. \ref{Fig:LayerLinks}, right). 
Since cells are described by the FitzHugh-Nagumo model, it is important to recall that the threshold is not involved in the single-neuron dynamics but it concerns presynaptic neurons at the synapse level. Indeed, when the presynaptic neuron exceeds the threshold, neurotransmitter release starts and influences the postsynpatic one.

\begin{figure}
\centering
\subfigure{\includegraphics[width=0.4\textwidth]{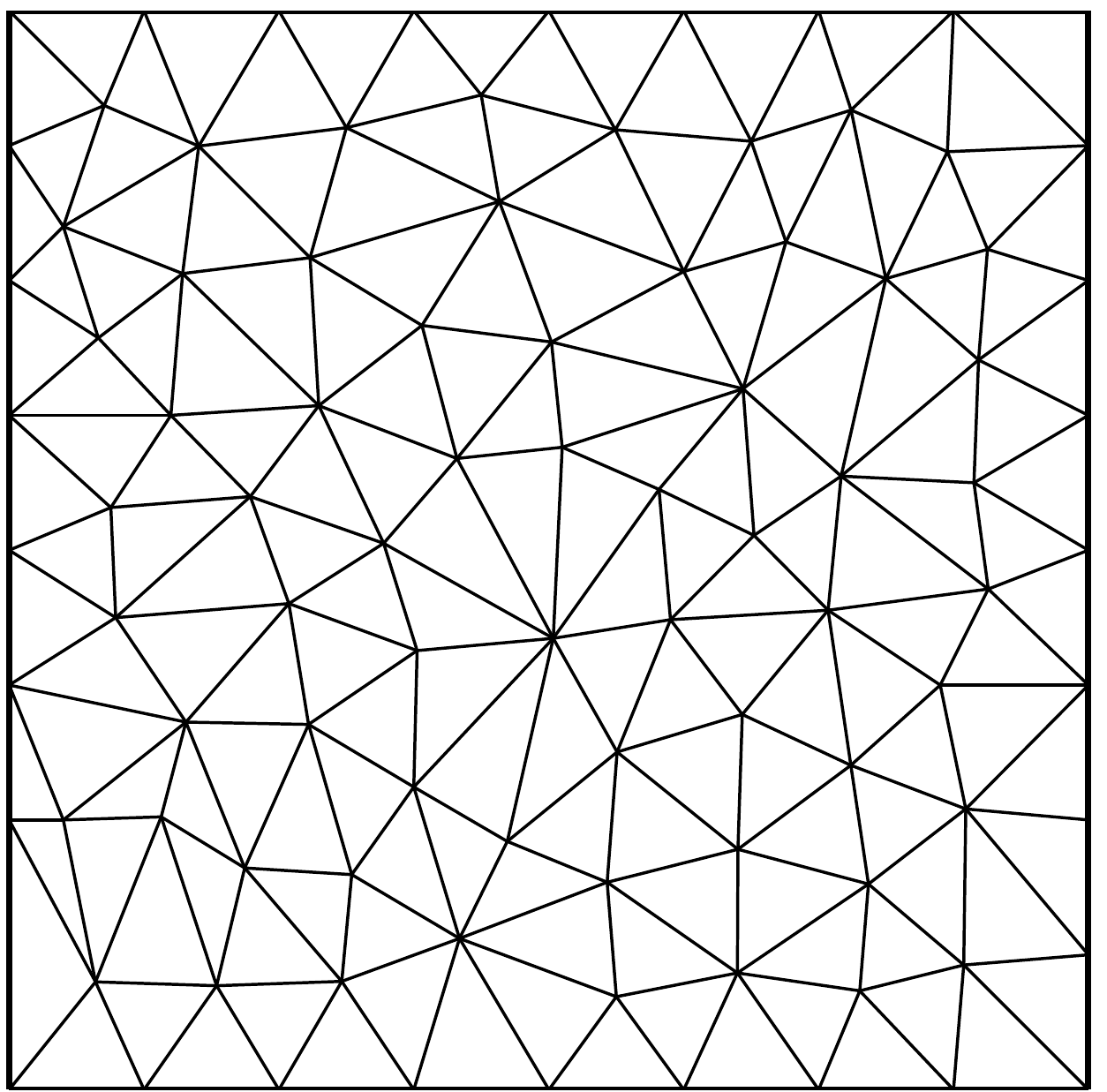}}
\caption{\label{Fig:MeshGrid}Domain decompositions obtained by exploiting the triangular mesh generator BBTR in \cite{BerroreBarbera}. The RefiningOptions parameter is set to 0.01, leading to a sparse mesh.}
\end{figure}

We close this section by a few words about the numerical treatment of our model. 
Concerning  GoCs - which form a discrete set - they are placed at the vertices of a quasi-uniform triangulation of the upper domain; we use the triangular mesh generator BBTR, described in \cite{BerroreBarbera}, with the mesh refinement parameter chosen to yield $250$ vertices (RefiningOptions parameter set to $0.0035$; Fig. \ref{Fig:MeshGrid}). On the other hand, GrCs - which form a continuum in our model - are described by a set of partial differential equations that need to be discretized in space. To this end, we resort to a classical second-order centered finite difference method (see, e.g. \cite{Quarteroni}). In particular, we consider 24000 nodes in the domain, lying on a regular grid, to represent the 300.000 GrCs. Therefore, using this grid size each vertex represents 12 or 13 GrCs. However, the results of the simulations turn out to be nearly independent on the GrCs grid refinement,  as it will be documented at the end of Section  \ref{Sec:CS}. At last, time integration of the resulting coupled system of ordinary differential equations is accomplished by the MATLAB routine ODE45. We remark that the spatial discretization might be accomplished by finite elements instead of finite differences, thus allowing for the easy use of unstructured grids that adapt themselves to the formation of localized patterns; this will be object of future work.

\section{Results and Discussion}
\subsection{Oscillatory activity in the granular layer} 
Numerical simulations have been performed with the aim to validate the capability of the hybrid network model, composed of  \eqref{Eq:MultiscaleDiscrete1}+\eqref{Eq:GoC} and \eqref{Eq:MultiscaleContinuum1}+\eqref{Eq:GrC}, to reproduce the GLN activity simulated in a biologically realistic model \cite{Solinas2010}. 
The network size is equivalent to a box with \unit{500}{\micro\meter} edge length along both the transverse and sagittal axes, and \unit{100}{\micro\meter} thickness containing the cubic volume (\unit{500}{\micro\meter} edge length) of brain tissue simulated in \cite{Solinas2010}.

Inspired by the orders of magnitude of the parameters in \cite{DeSchutter, Solinas2010}, we set:
\begin{equation}
g_{\rm{syn}}=1,\quad d=0.05,\quad I_{\rm{mossy}}^{i}=I_{\rm{mossy}}^{\rm{GoC}}=0.1\;,
\end{equation}
for the Golgi cell discrete model, and
\begin{equation}
\gamma_{\rm{syn}}=0.05,\quad \delta =0.005,\quad \mathcal{I}_{\rm{mossy}}(\xi)=\mathcal{I}_{\rm{mossy}}^{\rm{GrC}}=0.1\;,
\end{equation}
for the Granular cell continuous one. In particular, $\mathcal{I}_{\rm{mossy}}^{\rm{GrC}}$ is applied to $3\%$ of GrC nodes randomly chosen with uniform distribution. It is well known that  MFs input GoCs, as well as GrCs. Since in the real GLN also GoCs receive excitatory input from MFs. We assume that $3\%$ of GoCs receive $I_{\rm{mossy}}^{\rm{GoC}}=0.1$. The current is  applied for all $t>50$ ms. In the meanwhile, MF current is maintained active to $3\%$ of GrCs from $t>0$ ms. Both GrCs and GoCs which receive the external current are randomly chosen with uniform distribution. The thresholds $v_T$ and $\omega_T$ for GoCs and GrCs are both set to $0.5$.
The GoC potentials are described with vertical bars while GrC dynamics is shown with a continuous surface.

\begin{figure}
\centering
\subfigure{\includegraphics[width=0.27\textwidth]{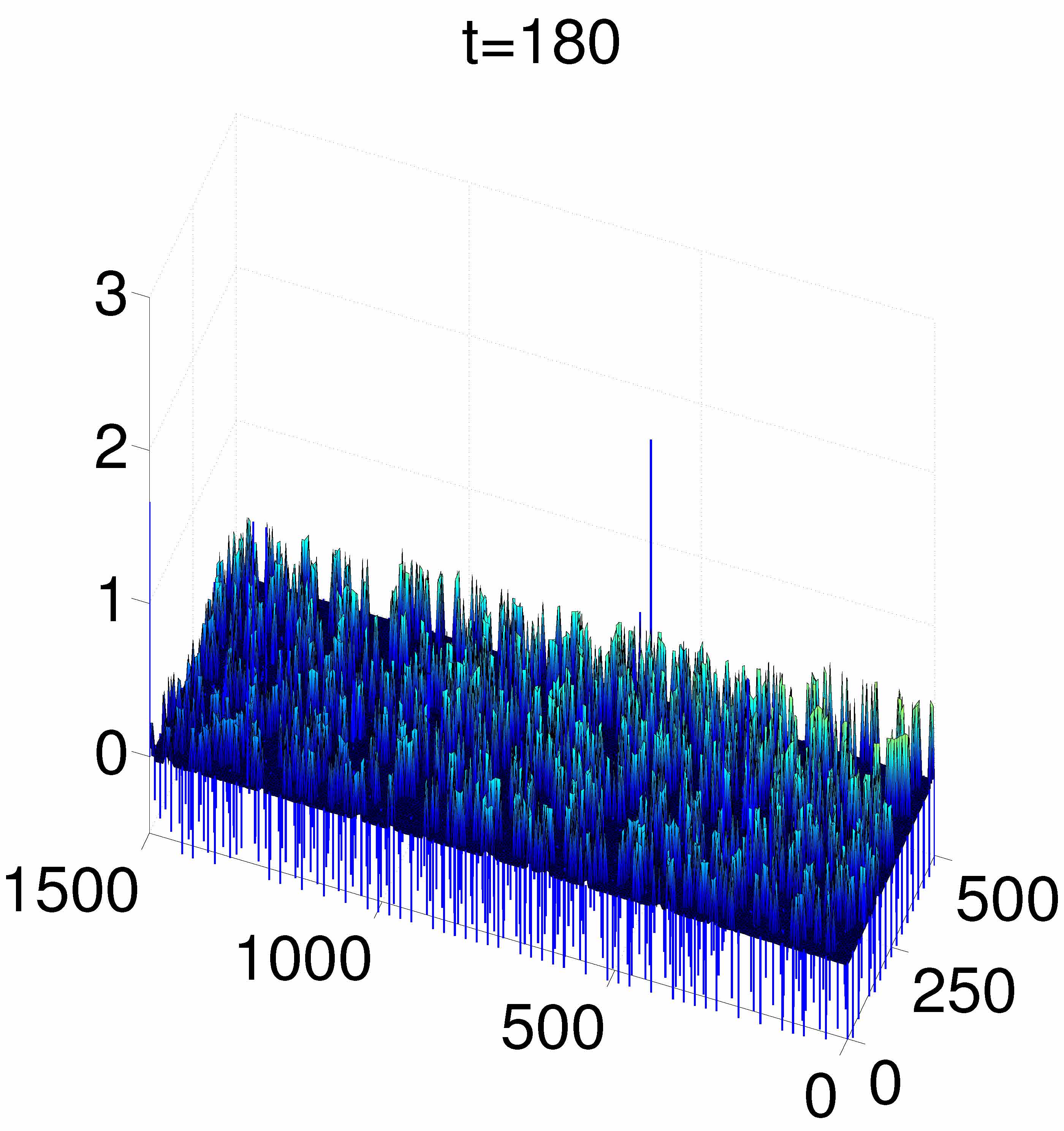}}
\subfigure{\includegraphics[width=0.27\textwidth]{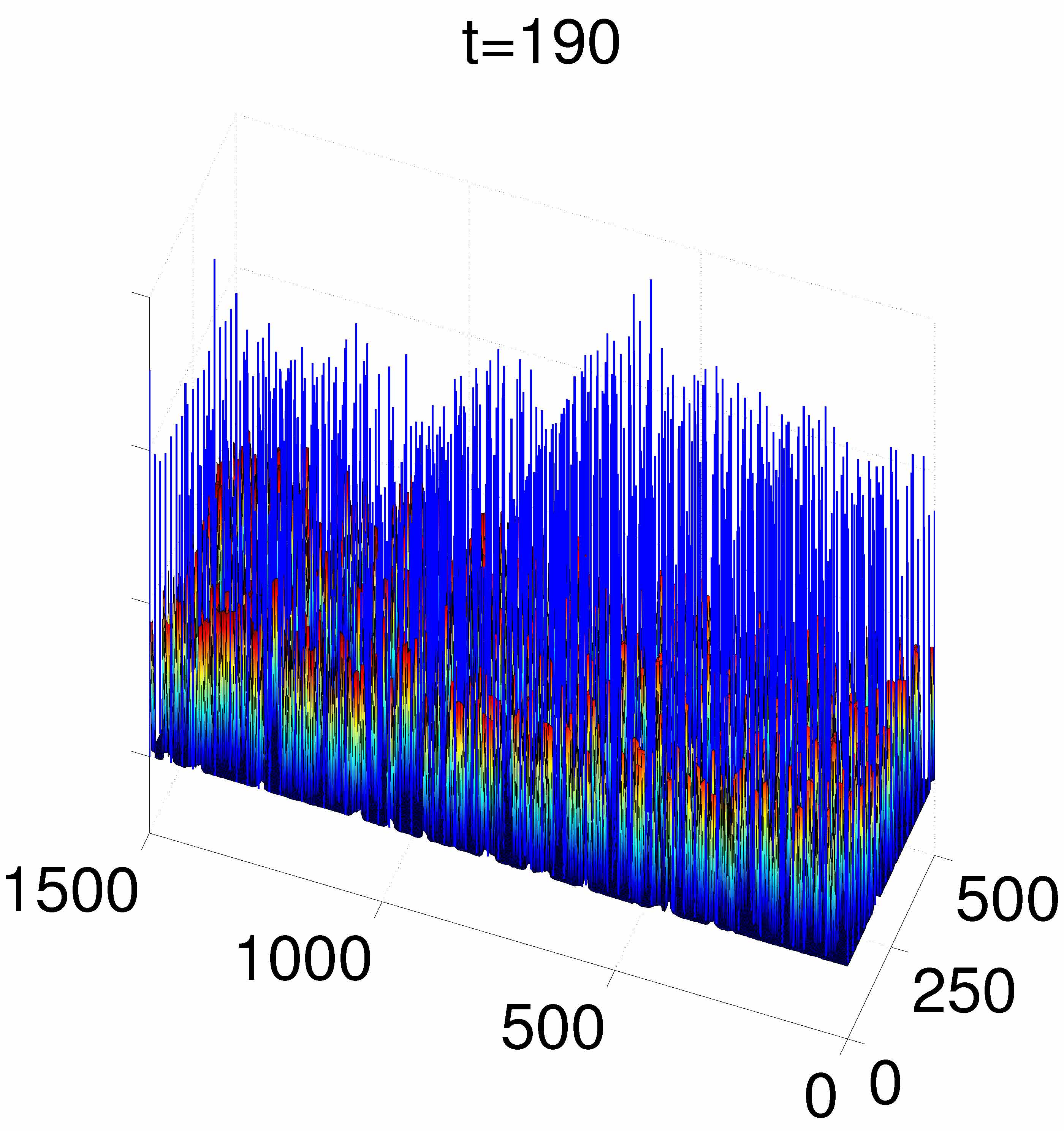}}
\subfigure{\includegraphics[width=0.35\textwidth]{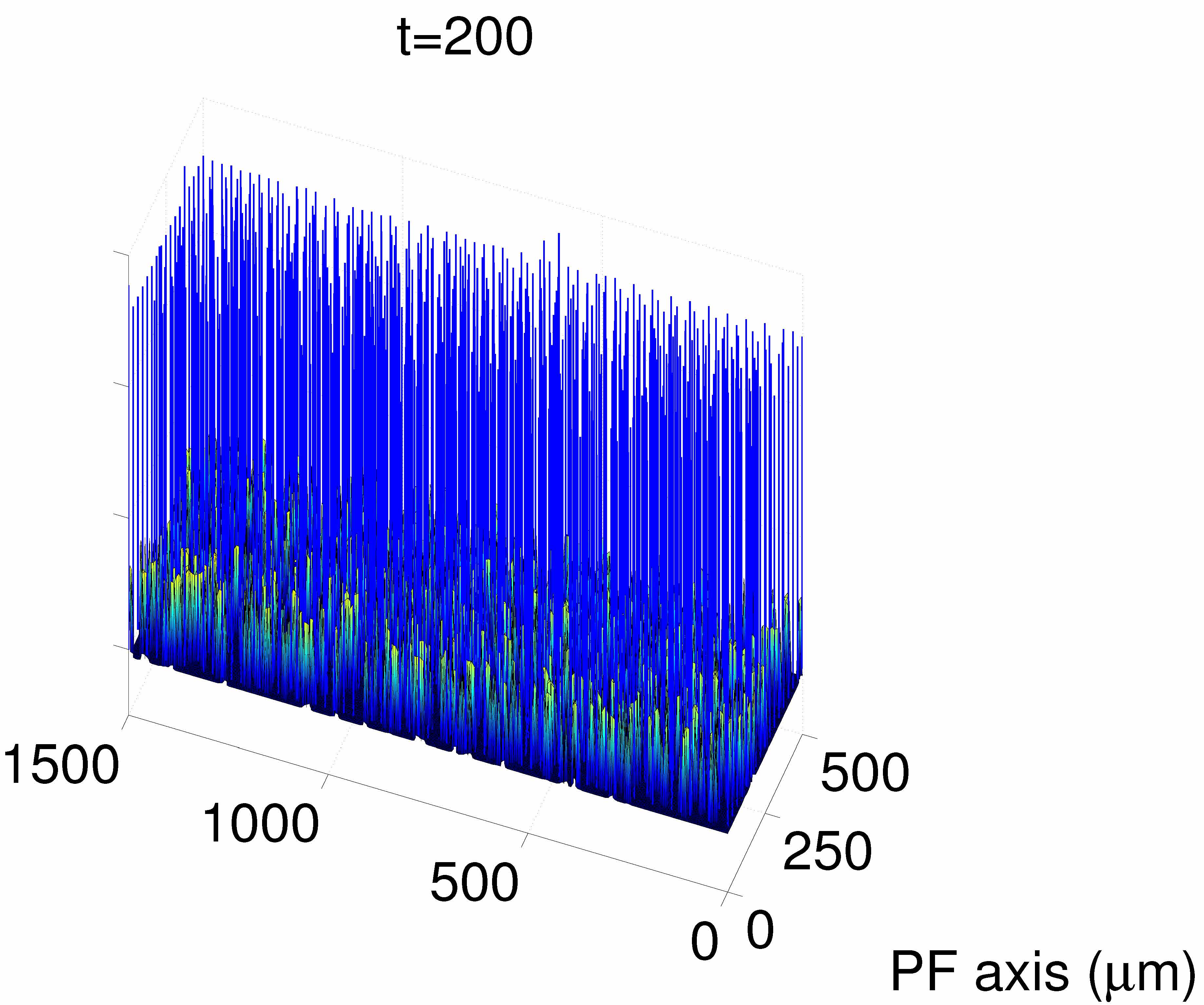}}\\\vspace{-0.30cm}
\subfigure{\includegraphics[width=0.27\textwidth]{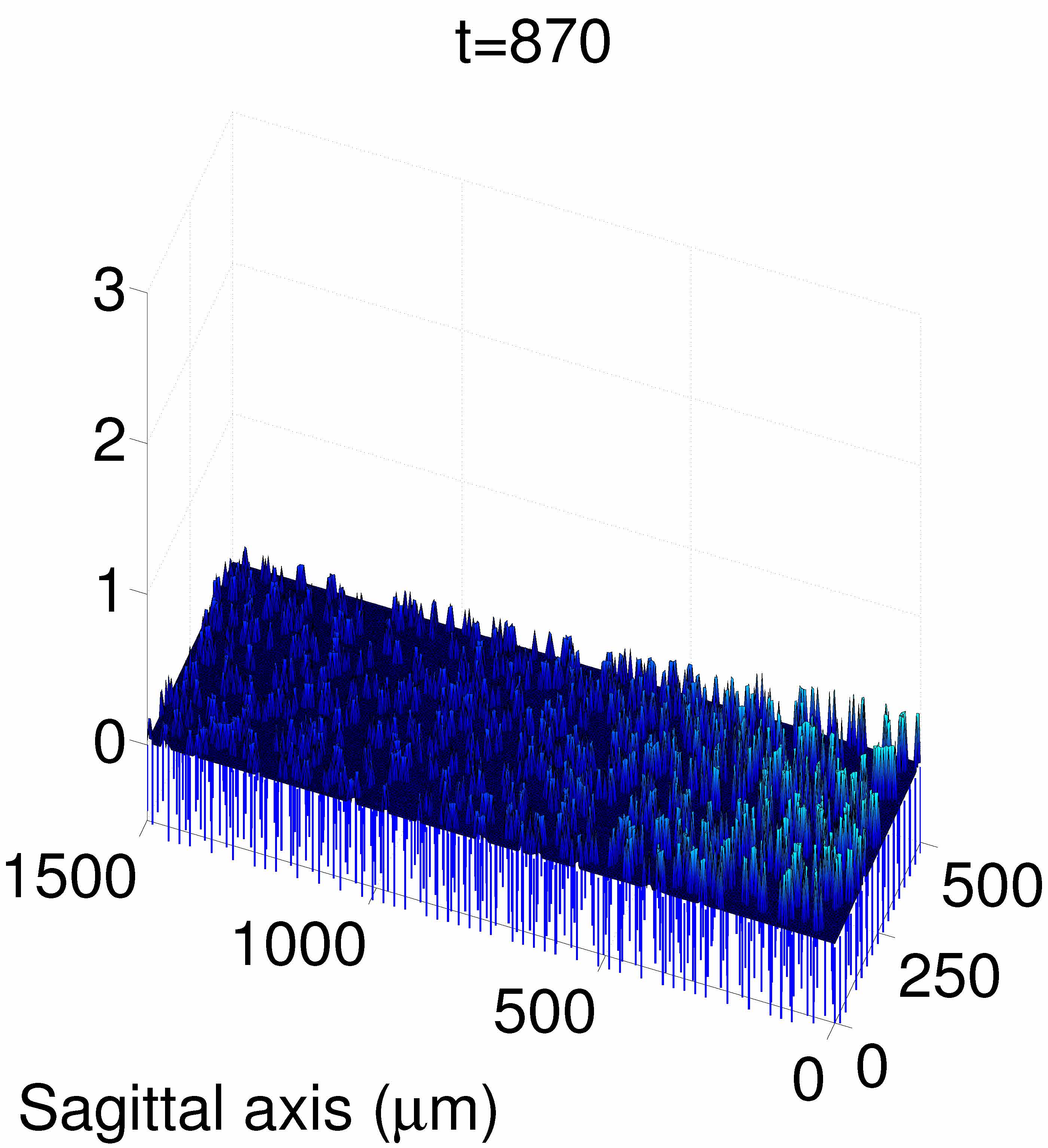}}
\subfigure{\includegraphics[width=0.27\textwidth]{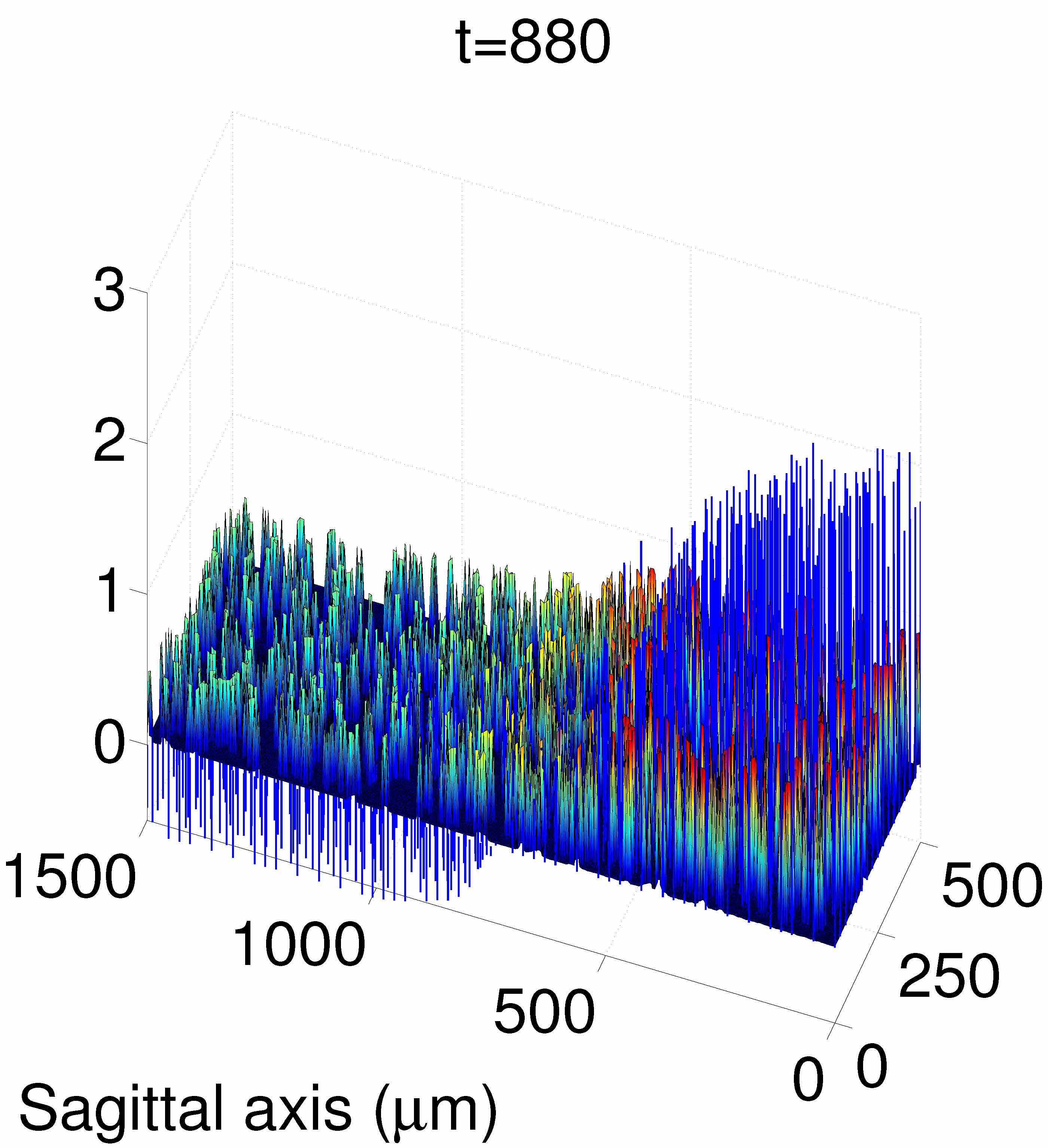}}
\subfigure{\includegraphics[width=0.35\textwidth]{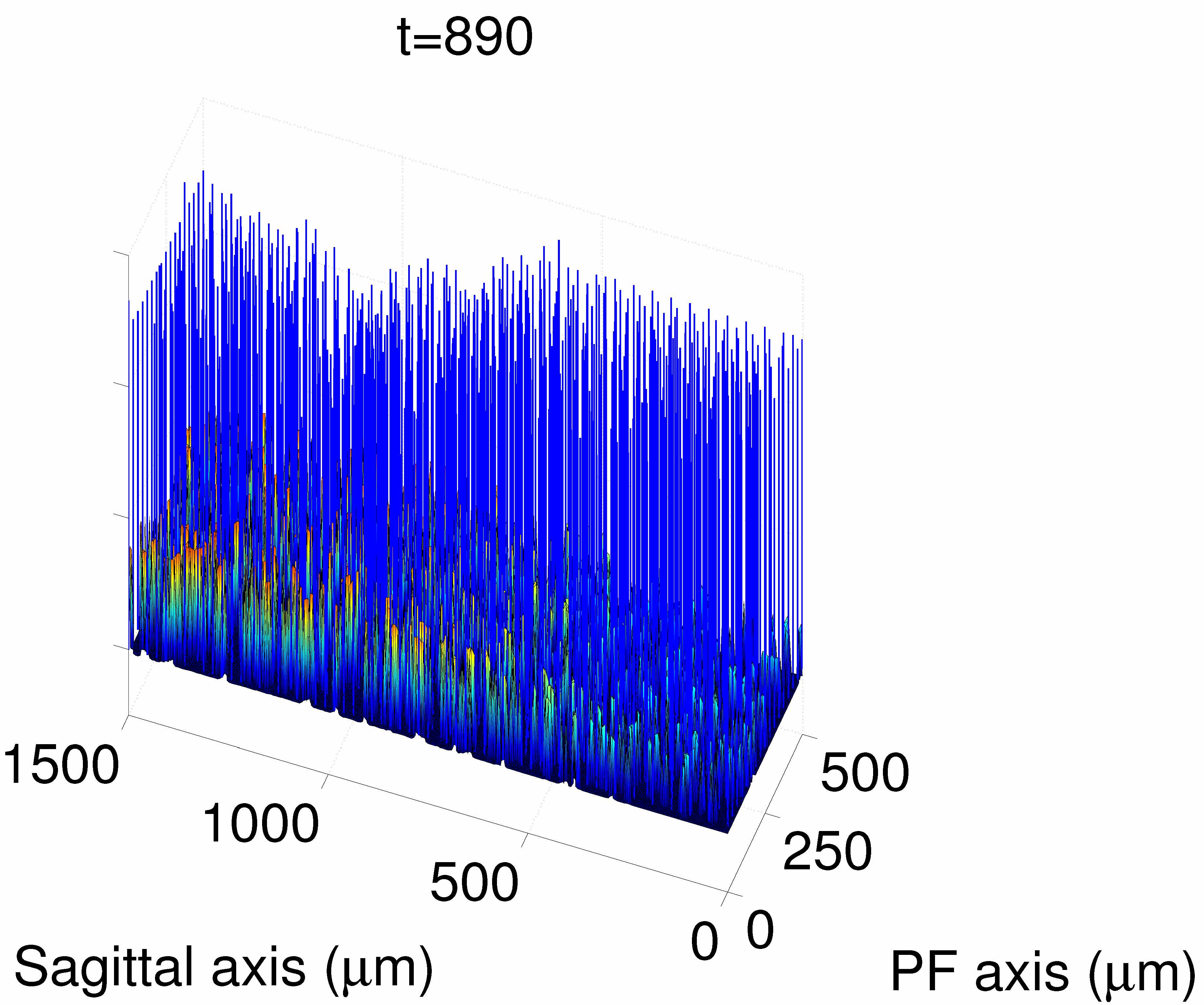}}
\caption{\label{Fig:MultiscaleMFtoGrAndGo}
Ensemble dynamics in the hybrid model. After an initial period of initialization ($t<180$ ms), a synchronous phenomenon within each population arises and the network activity shows oscillations with a frequency of $13$ Hz. After a few cycles ($t>350$ ms) a travelling wave phenomenono arises. The oscillatory frequency is unaffected by the spontaneous emergence of the waves. The waves of network activity diffuse along the sagittal axis. GrCs are represented with the coloured continuous graph; GoCs are described with bars showing potentials multiplied by a factor $3$ for graphical reasons. 
}
\end{figure}

A portrait of the GoC-GrC dynamics has been obtained by exploiting \eqref{Eq:MultiscaleDiscrete1}+\eqref{Eq:GoC} and \eqref{Eq:MultiscaleContinuum1}+\eqref{Eq:GrC}, assuming the topology described by \eqref{Eq:TopologyRect}. The excitatory input delivered by MFs to GrCs drives their activity above threshold and induces an increase in GoC potentials. The subsequent inhibition elicited in GrCs by the GoC inhibitory feedback loop (MF-GrCs-PFs-GoCs-GrCs) suppresses the GrC activity and the cycle restarts. 
The same local microcircuit synchronous phenomena do arise in the biologically realistic model of reference \cite{Solinas2010} and it is a characteristic dynamics observed in the GLN \textit{in vivo} \cite{Vos1999} and models \cite{Maex1998}. This dynamics is replicated with a specific period. Some significant snapshots are shown in Fig. \ref{Fig:MultiscaleMFtoGrAndGo}. Moreover, at a later time of the simulation, $t>400$ ms, the synchronous dynamics spontaneously converts in an interesting dynamics where excitatory waves travel in the whole domain involving both GoCs and GrCs, see Fig. \ref{Fig:MultiscaleMFtoGrAndGo}.

\subsection{Center-surround and time-windowing}
\label{Sec:CS}
Over the recent years several studies on the GoCs-GrCs network have been focused on the analysis of the integration of excitatory and inhibitory input by GrCs \cite{Mapelli2010combs,Mapelli2010a,Gandolfi2014,Nieus2014,Solinas2010}. 
To further validate our modelling reconstruction we focused on reproducing the spatial and temporal interaction of excitation and inhibition in the GLN following the work presented in \cite{Solinas2010}. 
According to \cite{Gandolfi2014,Solinas2010}, the input delivered by a small bundle of MFs in the GLN elicits the activation of a cluster of GrCs, a spot $33 \pm 5 \micro\meter$ wide at $70\%$ of the peak amplitude \cite{Mapelli2010combs}.
The spot is limited in size and in time by the properties of the feed-forward and feed-back inhibitory loops, due to the GoC integration properties and the arrangement of their axons. 
This phenomenon, defined center-surround and time-windowing in \cite{DAngelo2009}, is the result of the mismatch between the small area excited by the MFs and the wider area inhibited by GoCs activated directly and indirectly, through GrCs, by the same MFs, in combination with the inherent delay of the inhibitory loops.

This section is devoted to present noticeable center-surround and time-windowing phenomena reproduced by models \eqref{Eq:MultiscaleDiscrete1} and \eqref{Eq:MultiscaleContinuum1}. Furthermore, significant comparisons with results in \cite{Solinas2010} are presented. The aim of such comparisons is to show that our hybrid model reproduces the same dynamics shown in reference articles in the field. 
Now we set:
\begin{equation}
g_{\rm{syn}}=1,\quad d=0.005,\quad I_{\rm{mossy}}^{i}=I_{\rm{mossy}}^{\rm{GoC}}=0\;,
\end{equation}
for the Golgi cell discrete model, and
\begin{equation}
\gamma_{\rm{syn}}=0.05,\quad \delta =0.5,\quad \mathcal{I}_{\rm{mossy}}(\xi)=\mathcal{I}_{\rm{mossy}}^{\rm{GrC}}=0.2\;,
\end{equation}
for the Granular cell continuous one.

\begin{figure}
\centering
\subfigure{\includegraphics[width=0.25\textwidth]{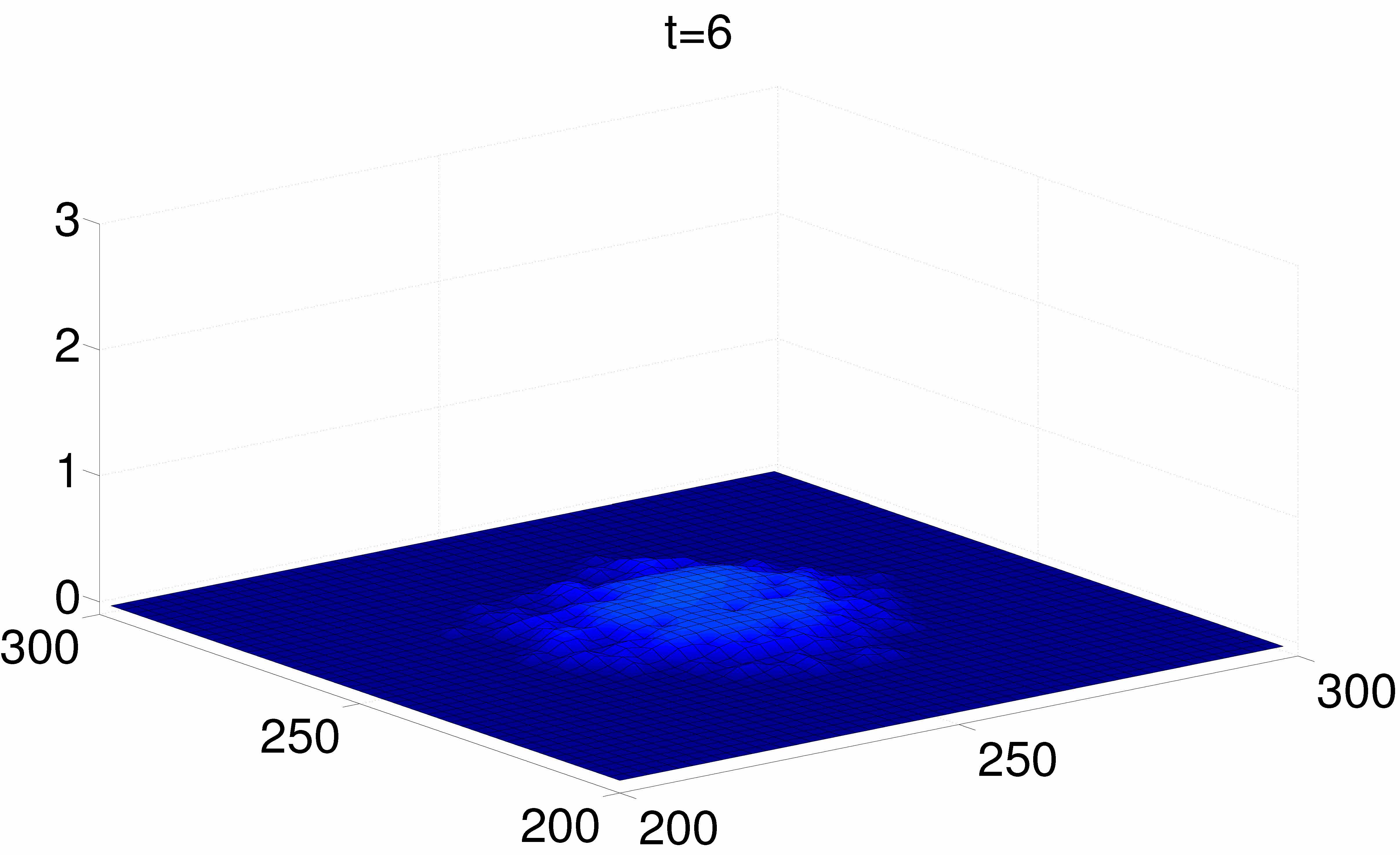}}
\subfigure{\includegraphics[width=0.25\textwidth]{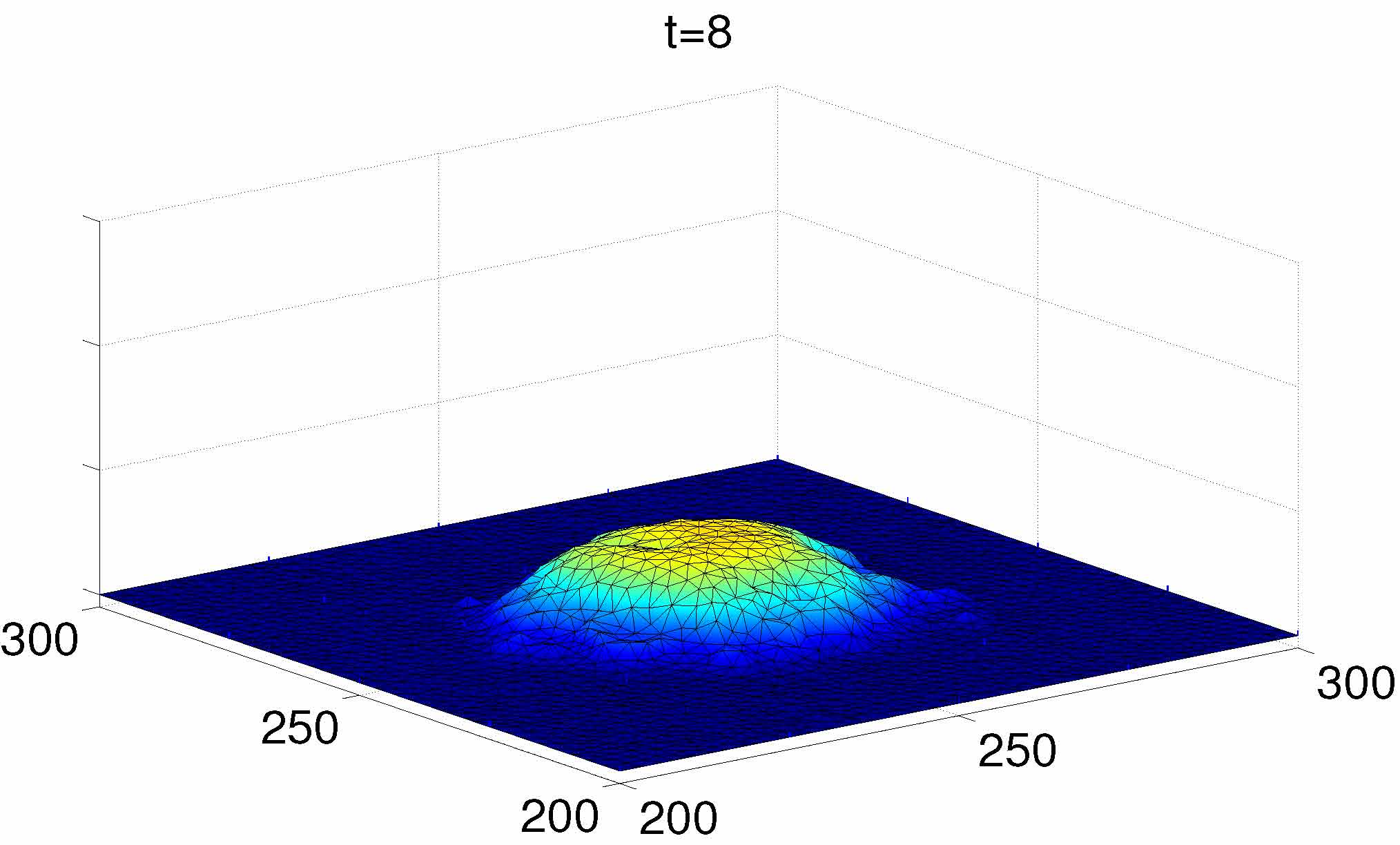}}
\subfigure{\includegraphics[width=0.25\textwidth]{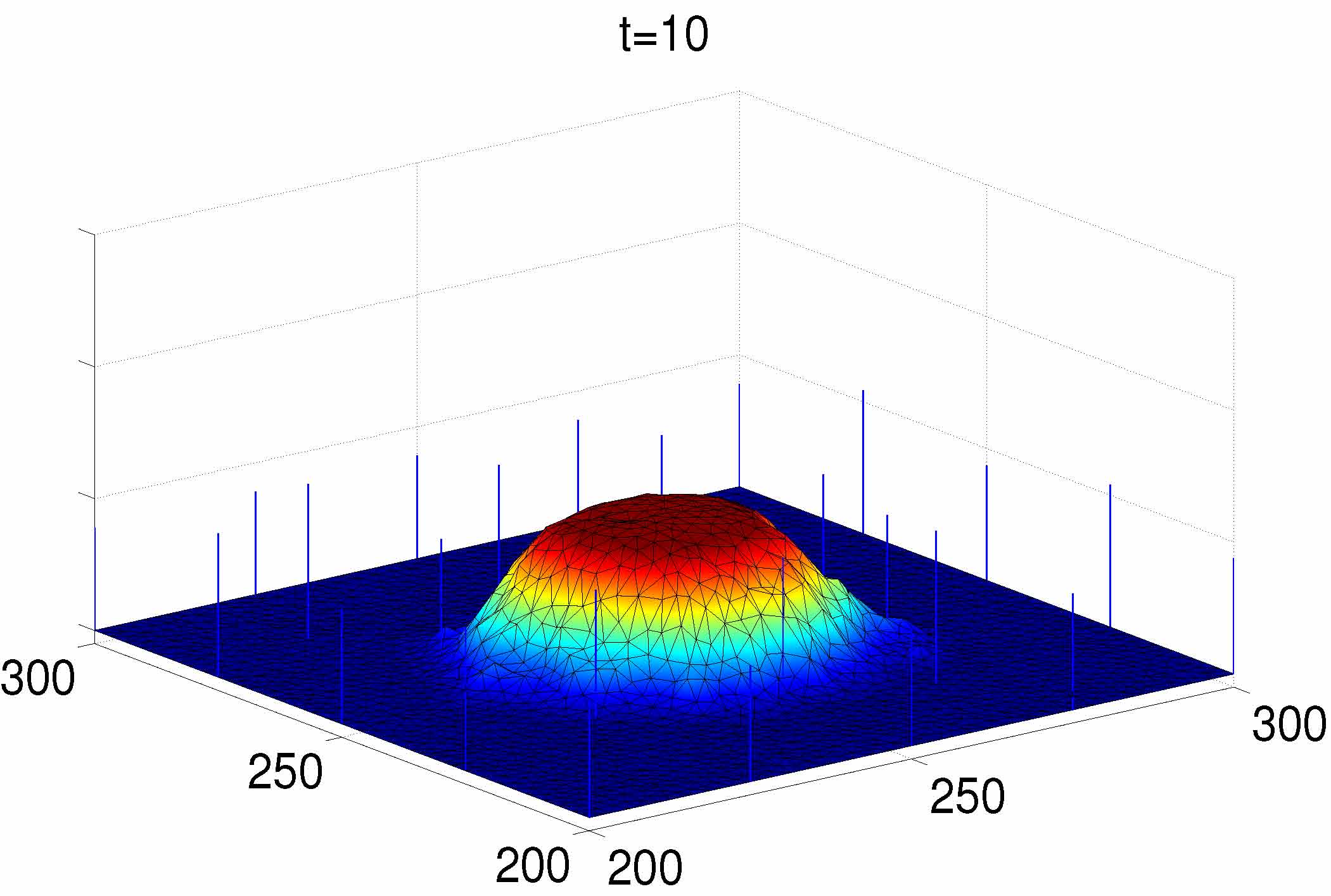}}\\
\subfigure{\includegraphics[width=0.25\textwidth]{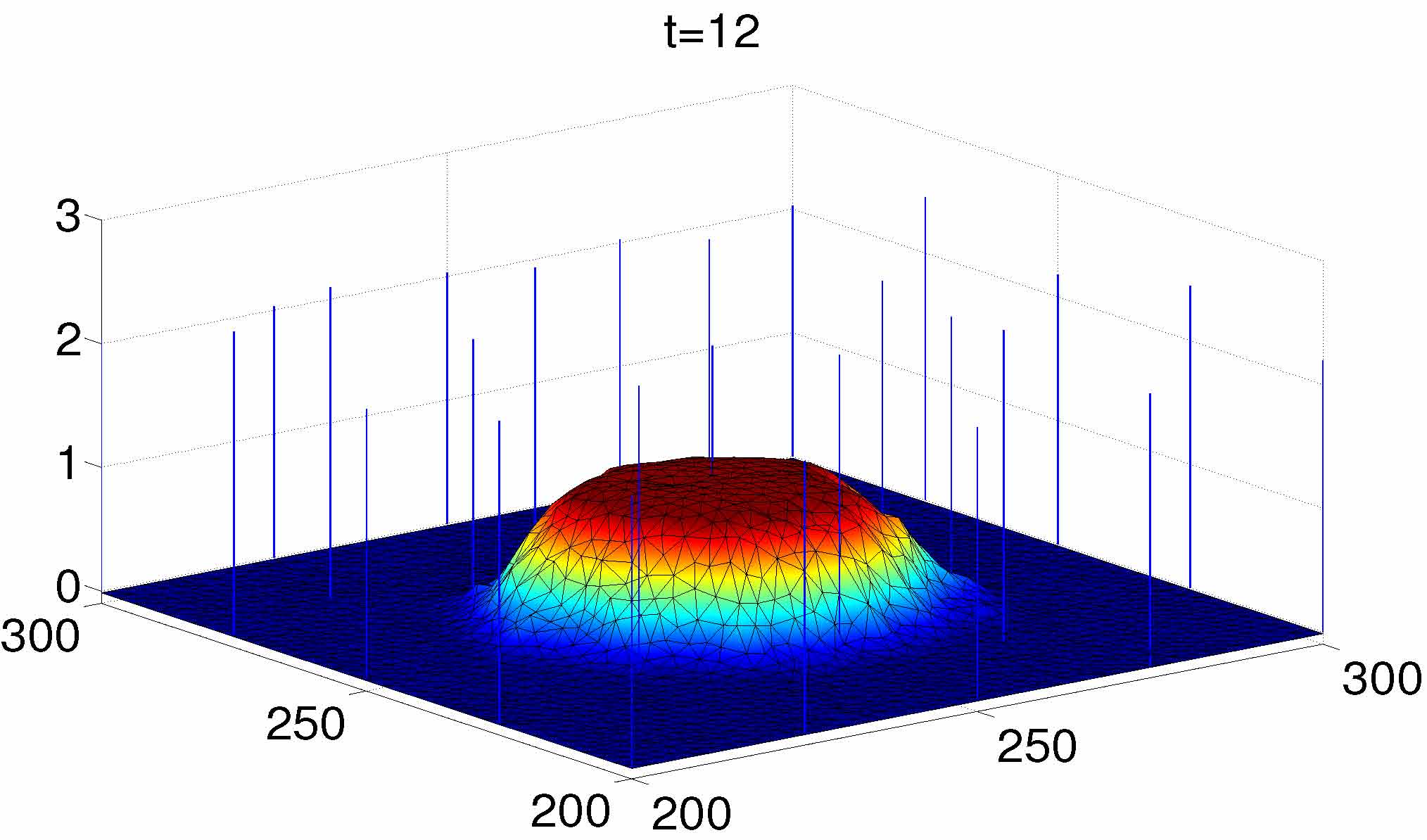}}
\subfigure{\includegraphics[width=0.25\textwidth]{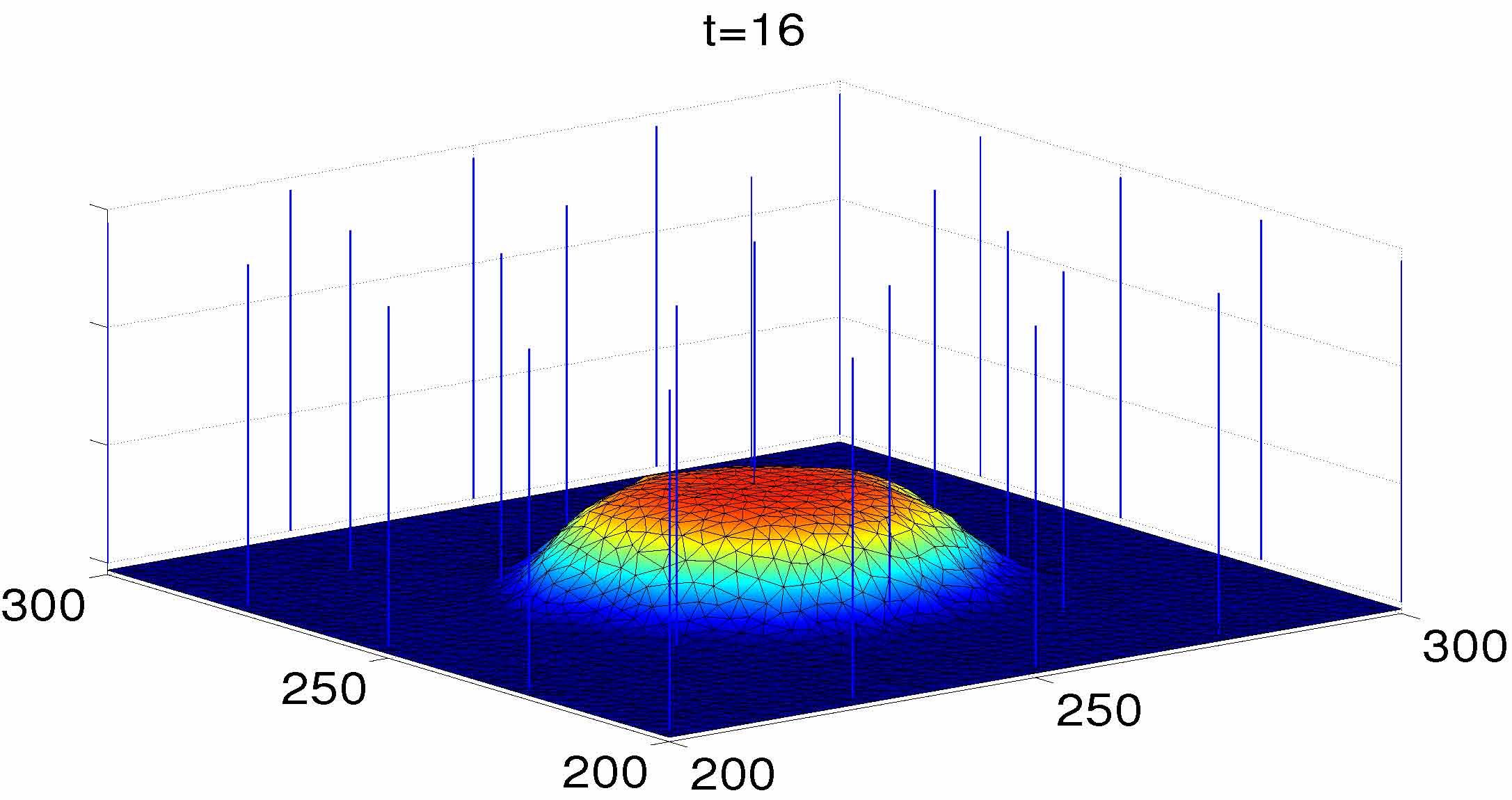}}
\subfigure{\includegraphics[width=0.25\textwidth]{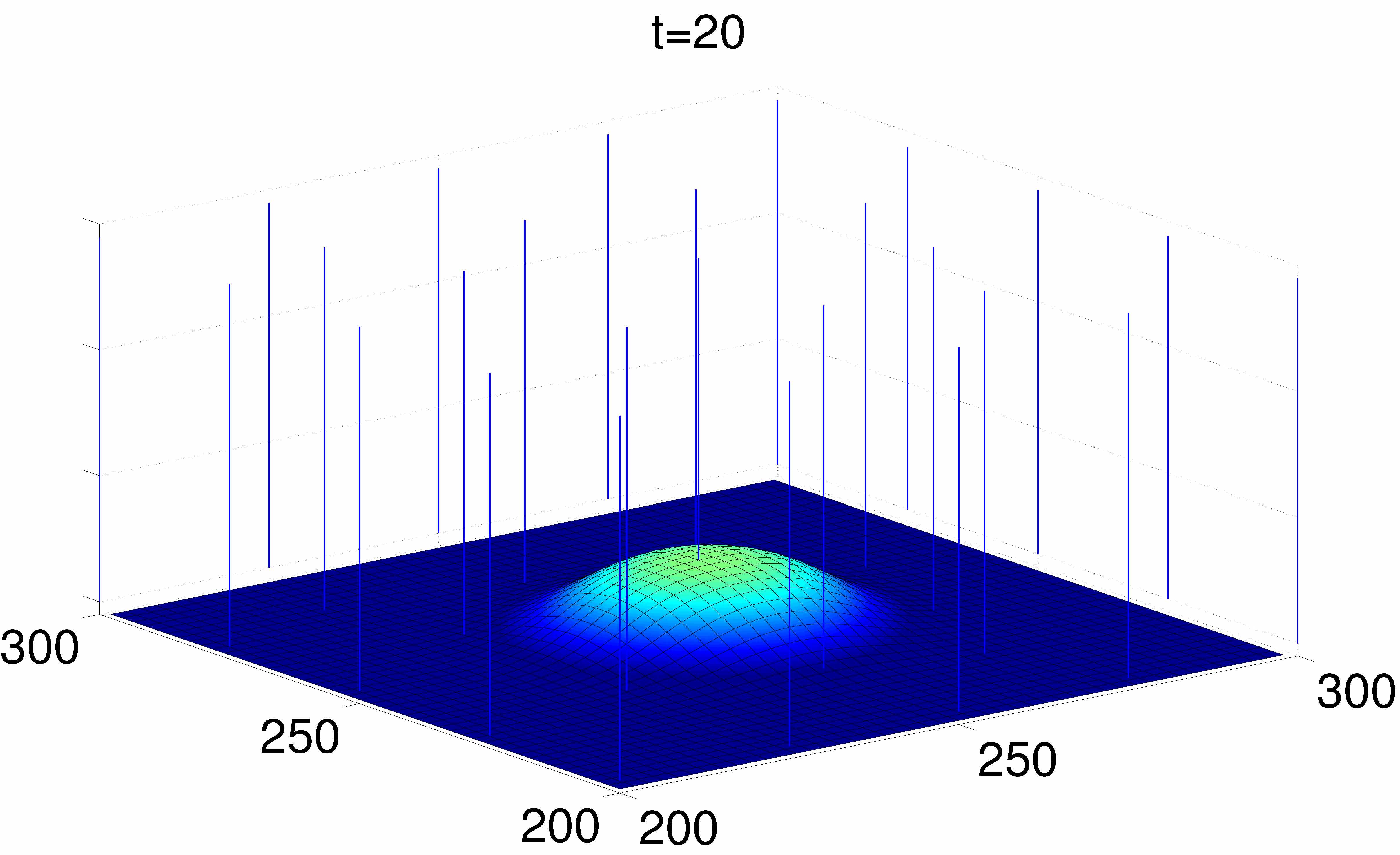}}\\
\subfigure{\includegraphics[width=0.25\textwidth]{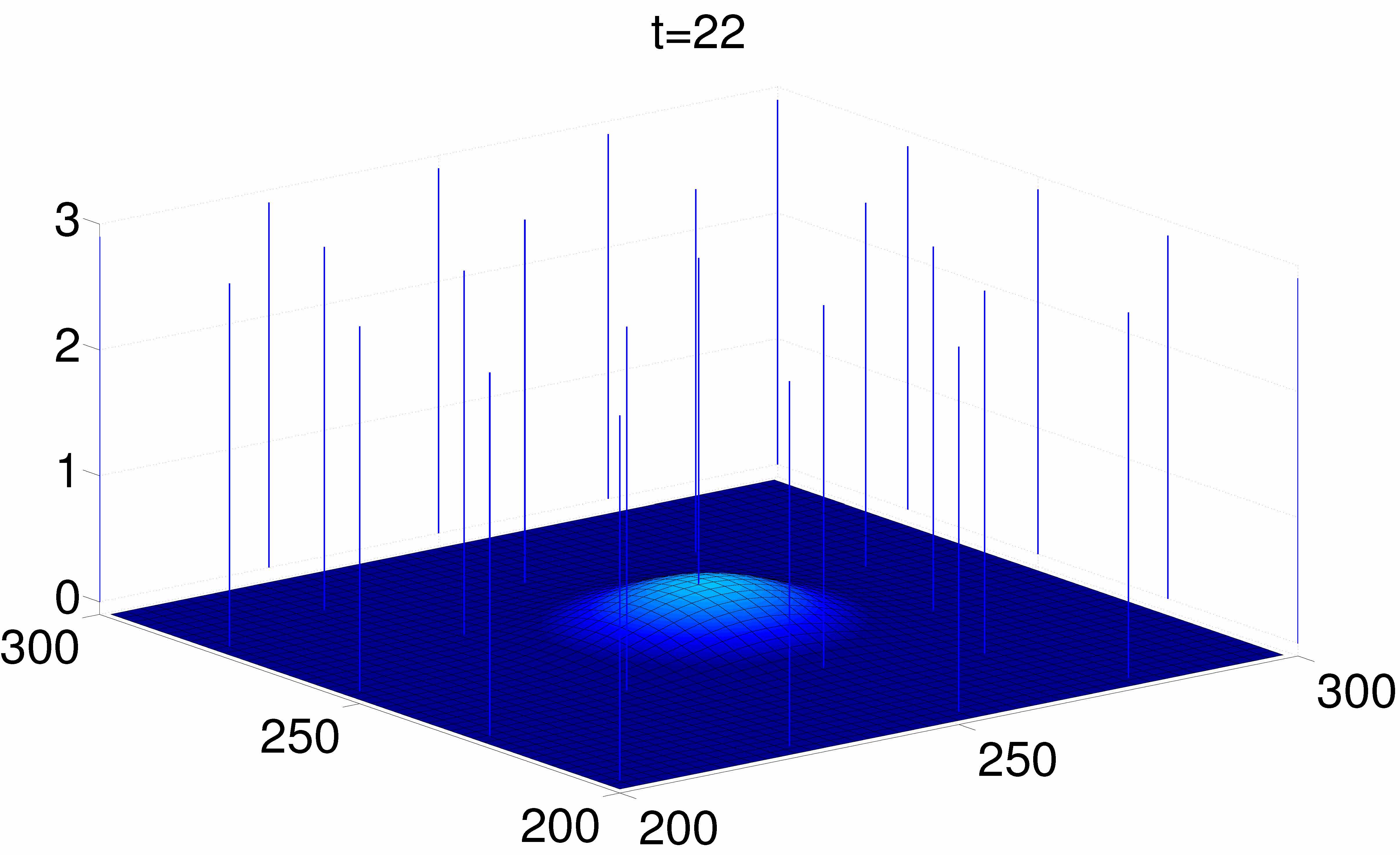}}
\subfigure{\includegraphics[width=0.25\textwidth]{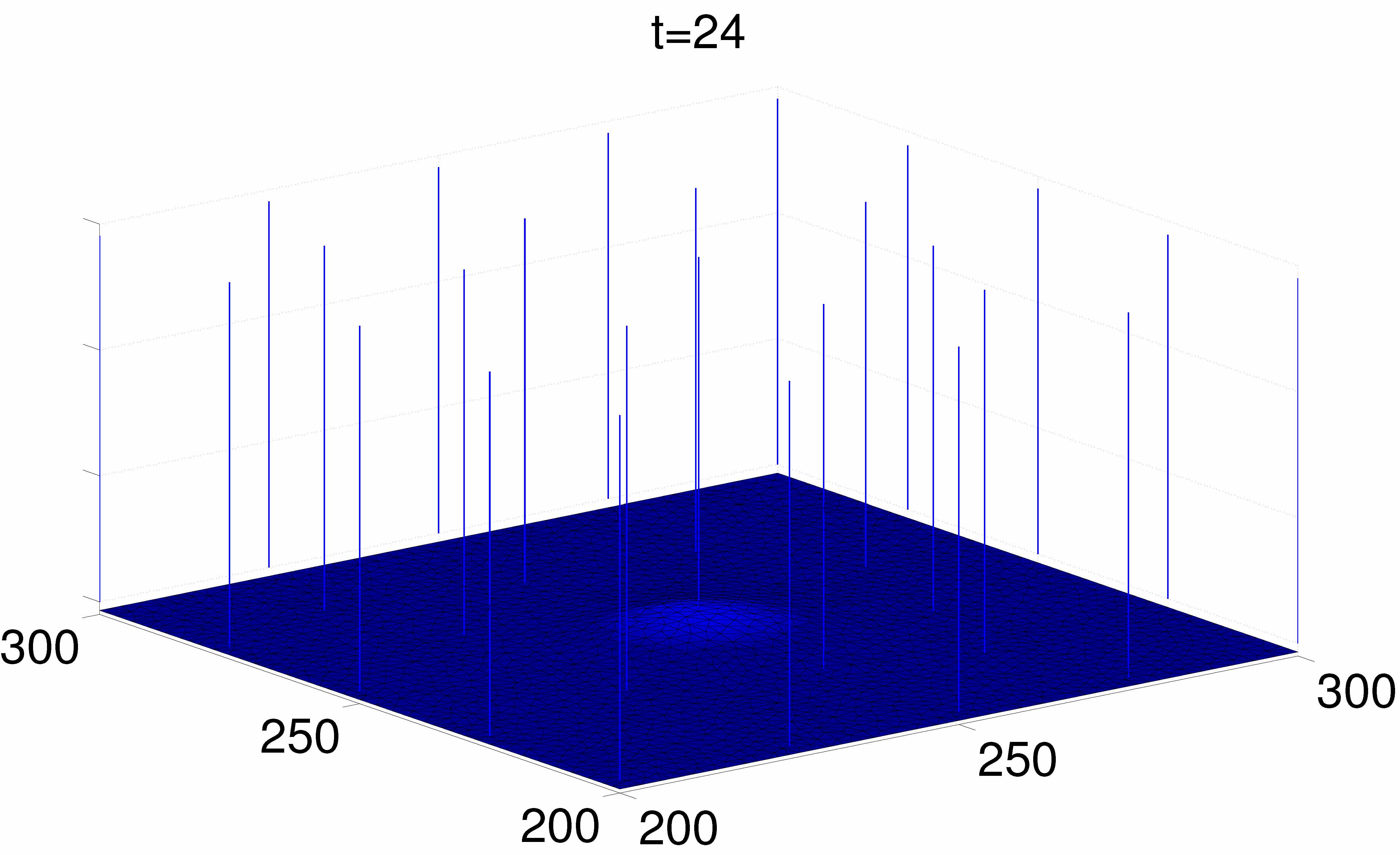}}
\caption{\label{Fig:CenterSurroundZoom}Snapshots describing center surround phenomenon. No excitatory inputs from MFs reach GoCs. On the contrary, GoCs are excited by GrCs through the PFs. In turn, each active GoC inhibits GrCs lying on a thin rectangle.
The stimulus was set on at $t=5$ and set off at $t=15$.
}
\end{figure}

The activation of a spot in the network centre was achieved in the original model by activating the MF terminals located within a sphere of radius equal to \unit{20}{\micro\meter} in the network centre. Considering that the average length of GrC dendrites was set to \unit{14}{\micro\meter} yielding an overall excited area of about \unit{34}{\micro\meter}. In the simulations we run to reproduce the impulse response of the GLN, we mimicked this activation by providing excitatory input to GrC vertices within a circle with radius equal to \unit{34}{\micro\meter}, $1/14.7$ in our normalised units, and located in the network centre. The simulation reproduced an activated spot in the network centre of the same size shown in \cite{Solinas2010} (data not shown). In a second simulation, we increased the activated area to \unit{50}{\micro\meter} in order to achieve a spot $33 \pm 5 \micro\meter$ wide at $70\%$ of the peak amplitude \cite{Mapelli2010combs} as shown in Figure \ref{Fig:CenterSurroundZoom} at $t=12$ ms. 



\begin{figure}
\centering
\includegraphics[width=1\textwidth]{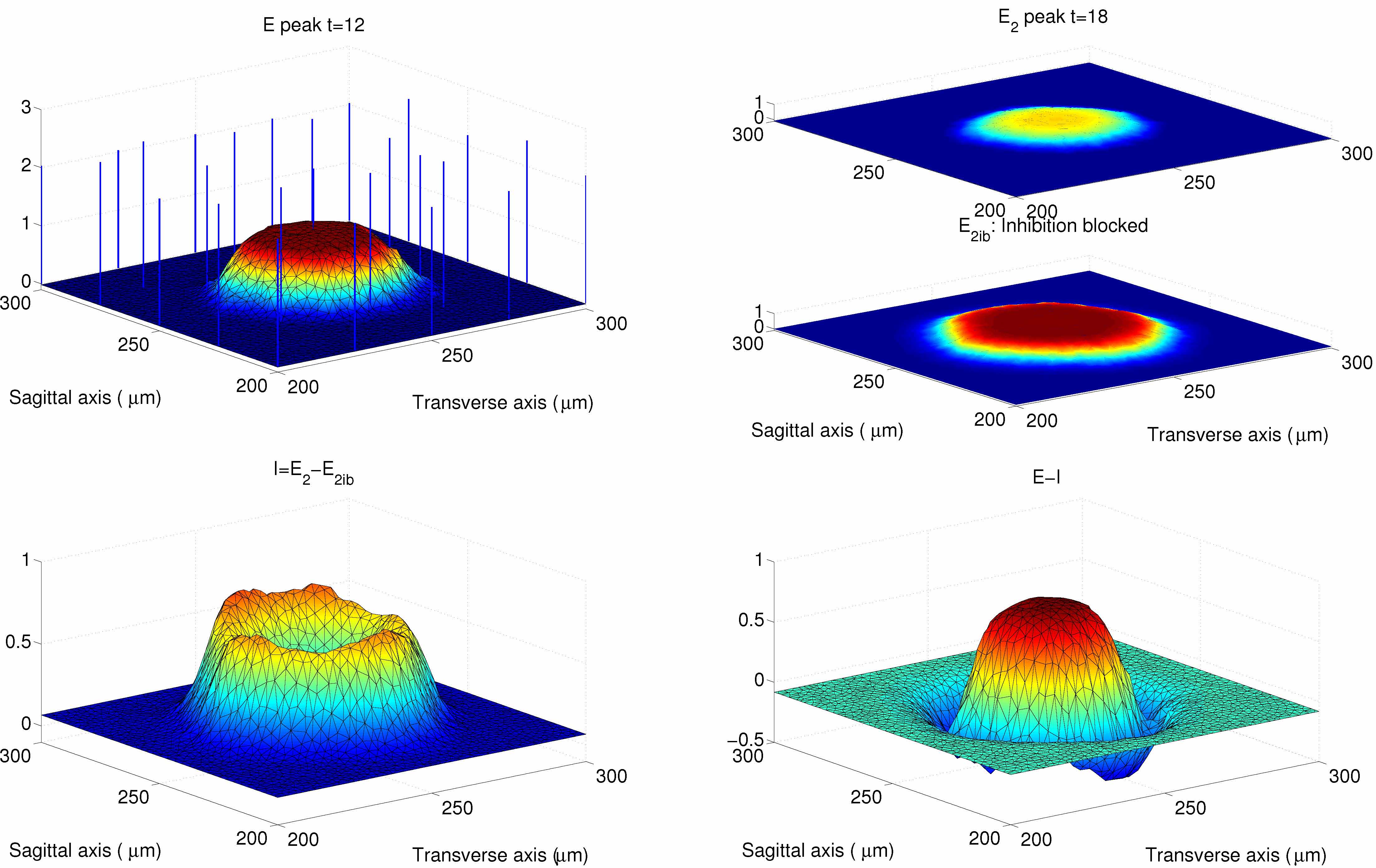}
\caption{\label{Fig:CompCS} The GLN was activated by a 10 ms pulse delivered by MFs to GrCs. The GLN activation at time 7 ms from the initiation of the stimulus shows the maximal activation yielded by the excitatory input (E peak; upper left panel). After 6 ms the GLN activation fades off due to the emergence of the inhibitory feedback ($E_2$ peak; upper right panel). After the block of inhibitory synapses the $E_2$ peak increses in amplitude and extension ($E_{\mbox{2ib}}$; upper right panel). The amount of inhibition is calculated as the change in GLN amplitude due to the block of inhibition at 13 ms from the stimulus initiation ($I$; lower left panel). The center-surround is represented as in \cite{Solinas2010} as the difference from the $E$ peak and the inhibition $I$ (lower right panel).
}
\end{figure}

Let us assume that the connection topology is again described by \eqref{Eq:TopologyRect} and let us consider MFs exciting GrCs in a circle, having radius $50$, located in the centre of the domain.
Figure \ref{Fig:CompCS} shows the GLN response to a stimulus set on at $t=5$ and set off at $t=15$ when the inhibitory connections were left active or blocked and their difference. The center-surround organisation of the inhibitory projections shapes the GLN response in space as it is evident from the enlargement of the active spot when those connections were blocked.

In order to compare our result with those shown in Fig. 5 of \cite{Solinas2010}, let us reproduce in Fig. \ref{Fig:CompCS} the same computational steps to evaluate the effect of inhibition of the GLN activation. 
After the onset of the MF input the GLN initiates its response with $1$ ms of delay reaching its maximal activation after $2$ ms, indicated as E peak in Fig. \ref{Fig:CompCS}. After 6 ms the GLN activation fades off due to the emergence of the inhibitory feedback and we choose this time to measure the $E_2$ peak. Blocking the inhibitory synapses the $E_2$ peak increases in amplitude and extension (inhibition blocked: $E_{\mbox{2ib}}$). As in \cite{Solinas2010}, the amount of inhibition $I$ is calculated as the change in GLN activity amplitude due to the block of inhibition. The center-surround is represented as in \cite{Solinas2010} as the difference from the $E$ peak and the inhibition $I$.


Let us recall that our model constituted by \eqref{Eq:MultiscaleDiscrete1} and \eqref{Eq:MultiscaleContinuum1} has been designed under strong simplifying assumptions that do not allow us to take into account the wide variety of phenomena in the single cell and in the whole network. Furthermore, the GrC layer has been described as a continuum. Nonetheless, the remarkable result obtained is that our model is able to reproduce the benchmark dynamics on the right-hand side, at least in the significant time range in which the center-surround phenomenon arises. Concurrently, the delayed activation of GoCs allows the response of GrCs to the stimuli to survive till the GoCs inhibition arises. This configures a time window where GrCs are allowed to transfer their activity to the subsequent network layers. The intervention of GoCs inhibition closes this window resetting the GrCs activity and making them ready to reliably transmit a new stimulus.

\begin{figure}
\centering
\subfigure{\includegraphics[width=0.49\textwidth]{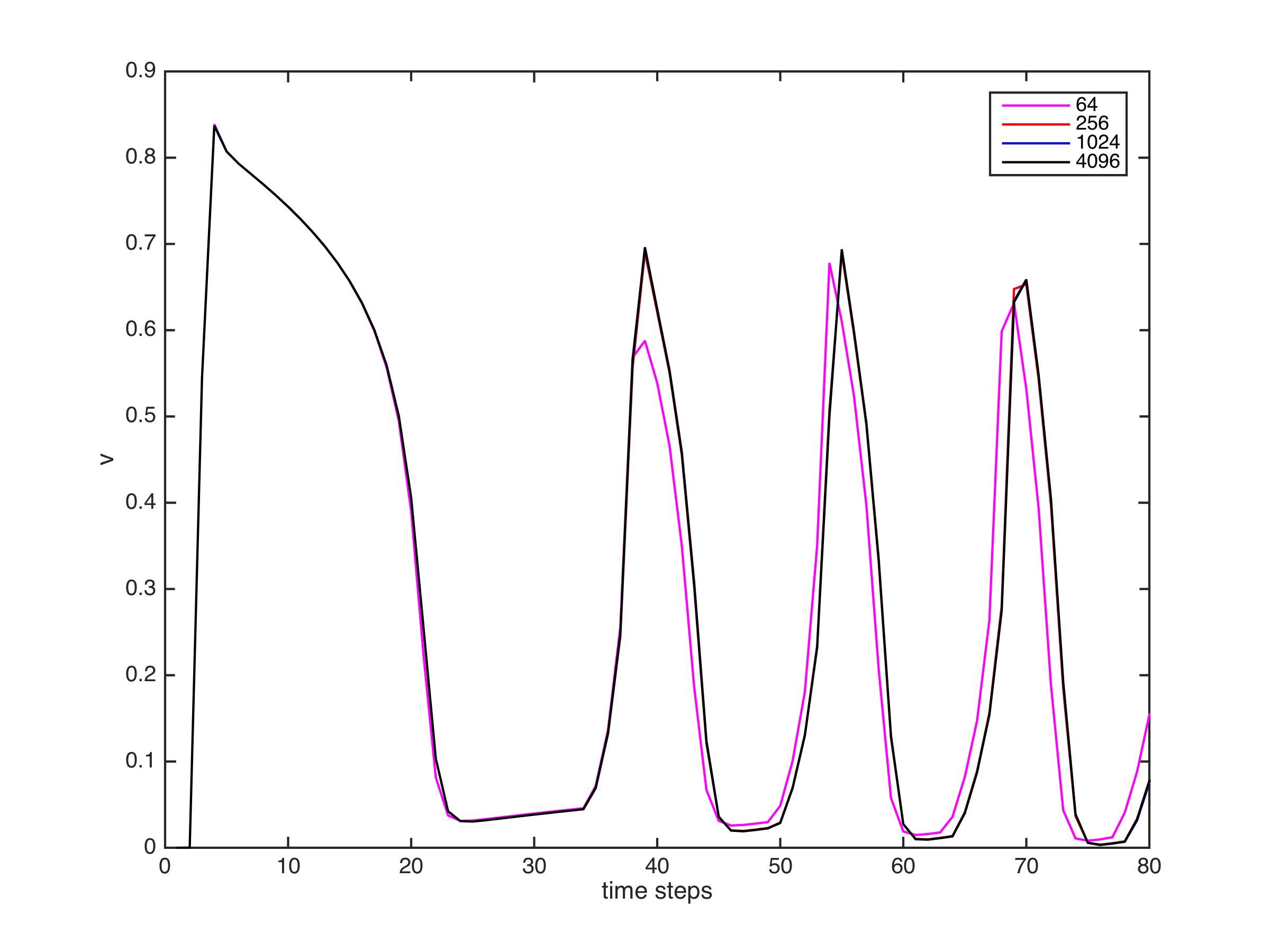}}
\subfigure{\includegraphics[width=0.49\textwidth]{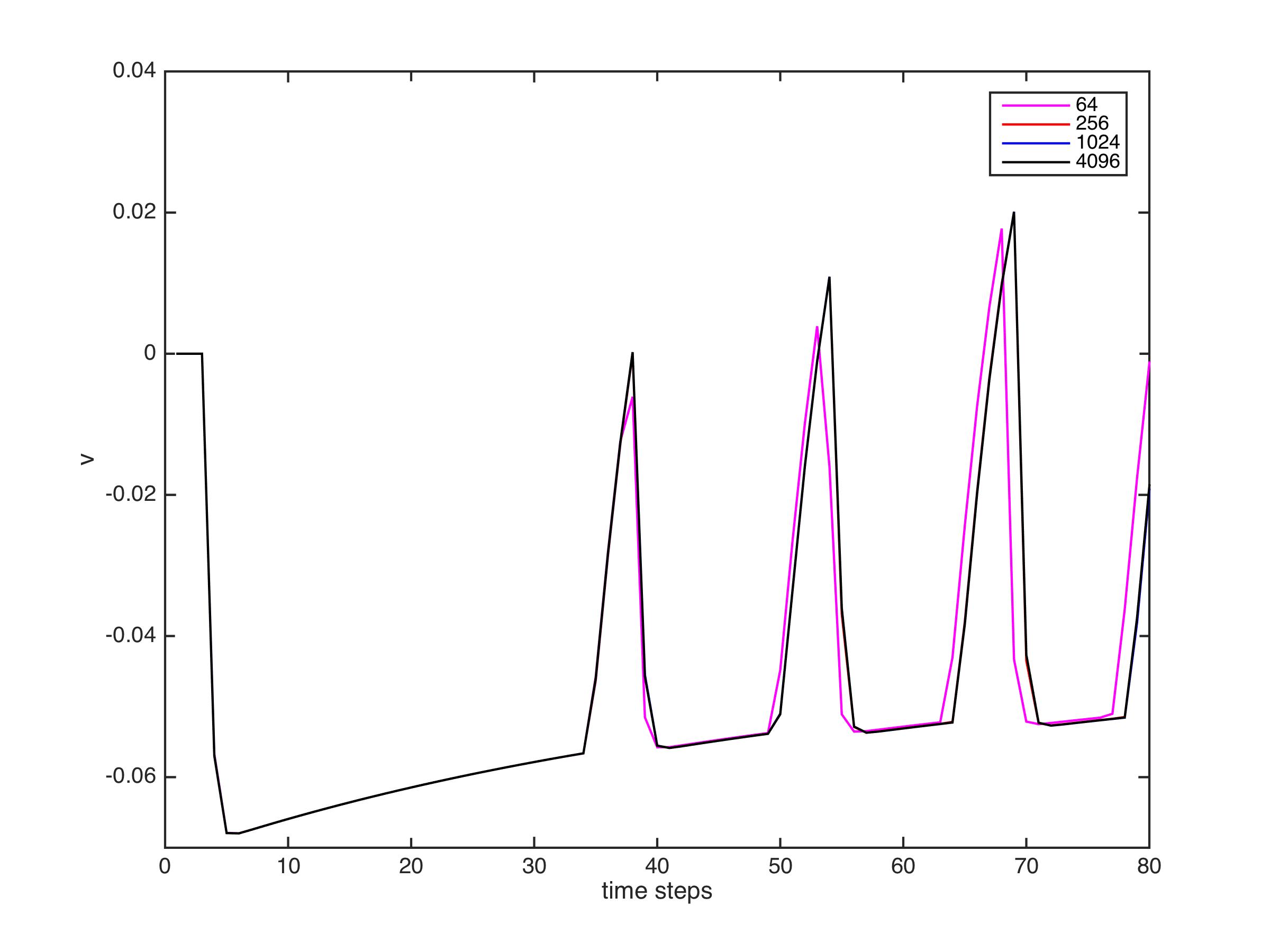}}
\caption{\label{Fig:Conv}Grid convergence for different structured grid resolutions of the continuous model for GrCs: 64, 256, 1024, 4096 nodes. The left panel shows the membrane potential $\omega$ plotted as a function of each time step for the node placed at $(0.5,0.5)$ in the centre of stimulation, i.e. subject to the excitatory input. The right panel shows the same plot for a node at $(0.4,0.4)$ outside the stimulated area and receiving only inhibitory input indirectly elicited by the feedback inhibitory loop.}
\end{figure}

Finally, we conclude the present section by stressing that the simulations provided in this paper turn out to be nearly independent on the GrCs continuous population grid refinement. Indeed, focusing on the framework that describes the center-surround phenomenon, we exhibit a comparison among the solutions produced by the model with increasing number of nodes in the space discretization of the GrC population. In Fig. \ref{Fig:Conv}, the evolution in time of the membrane potential of two cells in the domain is shown, for different values of the spatial resolution. Convergence is clearly documented, thereby providing a sound background to the use of our numerical simulator.

\subsection{Computational comparison}
The computational performance of our new modeling method was assessed by running a simulation with an equivalent representation of a portion of the GLN in both simulators: NEURON \cite{Neuron} and our hybrid model simulator. The simulation used as reference is the one reproducing the center-surround effect in \cite{Solinas2010}.
On the one hand, in \cite{Solinas2010} the full model simulation of 250 ms of network activity (simulation run using the code available at \cite{Solinas2010code} required about 428 sec on a Apple\r MacBook Pro (Intel Core 2 Duo $2.93$ GHz) for a network of 4001 GrCs and 27 GoCs. 
On the other hand, our network consists in $2116$ GrC vertices and $27$ GoCs. Considering an equivalent of 250 ms of activity, our simulation required about $71$ sec. Therefore, our network simulator is roughly 6.7 times faster than the NEURON simulator. We must also recall that the output of our simulator is immediately available for visualization in MATLAB while the output generated by the NEURON simulator requires an additional $30$ min of post processing to be visualized.
This analysis quantitatively confirms the reduced computational cost of employing our simplified model instead of a detailed one, without losing information about such fundamental activity in time and space as the center-surround and the time-windowing.
Let us stress that improvements of our codes will lead to further time simulation savings. The most significant one will consist in translating our routines into a programming language that could be compiled rather than interpreted, i.e. C rather than Matlab, and in restructuring our code in order to take advantage of the multithreading or parallelization programming feature of the C programming language.

\section{Conclusions}
With the aim of efficiently describing the dynamics of neuronal populations having a strong density difference in specific brain areas, the present work collects new results next to the ones presented in \cite{CanutoCattani}.
We started by stating the discrete conductance-based model \eqref{Eq:DiscCompleteModel} which describes the single cell membrane potential variation in time due to both electrical and chemical synapses. Afterwards, the derivation of the continuum model was obtained. By letting the number of neurons tend to infinity, we arrived at the complete model in \eqref{Eq:ContCompleteModel}.
The two models, discrete and continuous, were then coupled to describe populations exhibiting in specific areas of the brain significant differences in their densities, allowing us to formalize the hybrid model. Specifically, each cell of the low-density population was modelled by the discrete model, whereas the whole high-density population was described by the continuum model. Communications among populations, which translate into interactions among the discrete and the continuous models, are the essence of the hybrid model we presented. Such an approach, which may lead to a significant computational cost reduction, was applied to the Golgi-Granular network in the Cerebellum. Interesting dynamics such as synchronization, travelling waves, center-surround and time-windowing were reproduced by the hybrid model. The two latter dynamics were compared with recent results in literature devoted to this specific network, confirming the capability of our approach to reproduce significant dynamics.

By proceeding on the path here traced, some improvements should be taken into account in a forthcoming work. A major objective should concern how much the network behaviours here reproduced are related to the specific properties of the FitzHugh-Nagumo single-cell description. Moreover, one should evaluate if a different single-cell model is able to reproduce other significant behaviours such as resonant dynamics. Finally, in order to make the model more adherent to the reality, a future work should include the plasticity in communication strength among neurons.

\section*{Acknowledgements}
Part of this work has been done by the first author during the Ph.D. programme at the Polytechnic Institute of Turin.

We would like to thank Thierry Nieus and Diego Fasoli for enlightening discussions on various aspects of the present work.

The authors declare that there is no conflict of interest regarding the publication of this paper.

\bibliographystyle{plain}
\bibliography{Bibliography_MBE}
\end{document}